\documentclass[letterpaper]{article}

\pdfoutput=1
\usepackage[margin=1in]{geometry}
\usepackage{natbib}
\usepackage{hyperref}
\usepackage{despaper}
\usepackage{stmaryrd}
\newcommand*{\descpar}[1]{\noindent\textbf{#1:}\hspace{2ex}}
\usepackage{fontawesome5}
\newcommand*{\smilingimp}{\faUserSecret}

\newcommand*{\tss}{\textsubscript}
\newcommand*{\pout}{\tss{AC1}}
\newcommand*{\pin}{\tss{C2A}}

\title{Synthesis of Winning Attacks on Communication Protocols using Supervisory Control Theory: Two Case Studies}
\author{Shoma Matsui\thanks{Department of Electrial and Computer Engineering, Queen's University, Kingston, Canada. Email: \texttt{s.matsui@queensu.ca}}~~and~St\'{e}phane Lafortune\thanks{Department of Electrical Engineering and Computer Science, University of Michigan, Ann Arbor, USA. Email: \texttt{stephane@umich.edu}}}
\date{}

\begin{document}
    \maketitle

    \begin{abstract}
        There is an increasing need to study the vulnerability of communication protocols in distributed systems to malicious attacks that attempt to violate properties such as safety or nonblockingness. In this paper, we propose a common methodology for formal synthesis of successful attacks against two well-known protocols, the Alternating Bit Protocol (ABP) and the Transmission Control Protocol (TCP), where the attacker can \textit{always eventually} win, called \textsc{For-all} attacks. This extends previous work on the synthesis of \textsc{There-exists} attacks for TCP, where the attacker can \textit{sometimes} win. We model the ABP and TCP protocols and system architecture by finite-state automata and employ the supervisory control theory of discrete event systems to pose and solve the synthesis of \textsc{For-all} attacks, where the attacker has partial observability and controllability of the system events. We consider several scenarios of person-in-the-middle attacks against ABP and TCP and present the results of attack synthesis using our methodology for each case.
    \end{abstract}


    \noindent {\small\textbf{Statements and Declarations:~}The authors declare that they have no conflict of interest.}

    \section{Introduction}\label{sec:introduction}

Keeping systems secure against attacks and preventing security incidents are challenging tasks due to the increasing complexity of modern system architectures, where a number of hardware and software components communicate over potentially heterogenous networks. To analyze systems which are too complex to be fully described monolithically, abstraction employing formal methods plays a key role and it has been studied in particular in the computer science literature (see, e.g.,~\cite{baier_principles_2008,kang_multi-representational_2016}). In networked systems, components with different architectures cooperate with each other using various pre-designed \textit{protocols}. Due to the proliferation of communication using standardized protocols, vulnerabilities or misuses of protocols can result in serious security issues. As a concrete example,~\cite{bjorner_detection_2015}~introduces a formal model and analysis of a protocol used in Android OS, one of the most popular operating systems for smart phones. In order for components to cooperate with each other without damaging systems and without data corruption, robustness of protocols against communication failures is essential in modern system architectures. To ensure such robustness of protocols, relevant properties, such as \emph{safety} and \emph{liveness}, should be satisfied even if packets are dropped for instance. However, the situation is different in the context of malicious attacks, where an attacker that has infiltrated part of the system (e.g., the network) may be able to induce a violation of the safety or liveness properties, thereby causing the protocol to enter an abnormal state.

The development of resilient protocols that satisfy requirements and are applicable to various systems requires formal methods for modelling, verification, and synthesis. These problems have a long history in computer science as well as in control engineering. The readers are referred to \cite{baier_principles_2008}~and~\cite{holzmann_design_1991} for a comprehensive treatment of modelling and verification by employing formal methods, such as temporal logic. To prevent systems from being damaged by attacks that exploit vulnerabilities of protocols, the recent work \citep{Alur2017}~introduces the process of completing an incompletely specified protocol so that the completed protocol satisfies required properties and does not suffer from deadlock. \cite{Alur2017}~explains its methodology of protocol completion using the Alternating Bit Protocol (ABP).

In control engineering, the formalism of of discrete event systems (DES)~\citep{Cassandras2008} and its supervisory control theory (SCT)~\citep{wonham_supervisory_2019} are useful tools to treat the problem of protocol verification as a supervisory control problem~\citep{rudie_protocol_1992}, so as to determine whether a given protocol satisfies the required properties. Not only can SCT be used to analyze existing protocols, it can also be used to synthesize a desired protocol based on given requirements. For instance, \cite{kumar_discrete_1997}~introduces a systematic approach to design a protocol converter for mismatched protocols so that the specifications of the entire system and protocols themselves are satisfied simultaneously. On the other hand, \cite{rudie_communicating_1990}~considers protocols comprising local communicating processes, and formalizes protocol synthesis as the problem of controlling the local processes so that the global specification of the entire system is satisfied, employing the decentralized version of SCT. For a comprehensive survey of protocol synthesis, focusing on the formalization of the design of protocols, the readers are referred to~\cite{saleh1996synthesis}.

More generally, detection, mitigation, and prevention of attacks on supervisory control systems within the framework of SCT has been considered in several works, such as~\cite{carvalho_detection_2018,wakaiki_supervisory_2019,su_supervisor_2018,meira-goes_towards_2019}. \cite{carvalho_detection_2018}~presents a methodology of designing intrusion detectors to mitigate online four types of attacks; actuator enablement/disablement and sensor erasure/insertion. Focusing on sensor deception attacks under which the attacker arbitrarily edits sensor readings by intervening between the target system and its control module to trick the supervisor to issue improper control commands, \cite{wakaiki_supervisory_2019}~and~\cite{su_supervisor_2018}~study how to synthesize robust supervisors against sensor deception attacks, while \cite{su_supervisor_2018}~also introduces the synthesis problem of attack strategies from the attacker's point of view. Subsequently, a different technique from \cite{su_supervisor_2018} to compute a solution of the synthesis problem of robust supervisors was proposed in \cite{meira-goes_towards_2019}.

As protection against attacks is one of the main subjects of systems security, methodologies for designing attack strategies against systems have been reported in the literature~\citep{meira-goes_synthesis_2020,lin_synthesis_2019,VonHippel2020}. \cite{meira-goes_synthesis_2020}~presents how to synthesize an attacker in the context of stealthy deception attacks, modelled in the framework of SCT, which cannot be detected by the supervisor and cause damage to the system, as a counter weapon against intrusion detection modules as in~\cite{carvalho_detection_2018}. While \cite{meira-goes_synthesis_2020} considers sensor deception attacks as the attacker's weapon, \cite{lin_synthesis_2019} introduces the synthesis of actuator attacks under which the attacker has the ability to hijack the control commands generated by the supervisor, to damage the system.

Formal synthesis of successful attacks against protocols is the problem considered in this paper, in the context of two case studies. The work in \cite{VonHippel2020} (and its conference version~\citep{casimiro_automated_2020}) is of special relevance, as it introduces a methodology of attacker synthesis against systems whose components are modelled as finite-state automata (FSA). It presents how so-called ``\textsc{There-exists}'' attackers can be found (if they exist) using a formal methodology that has been implemented in the software tool \textsc{Korg}~\citep{korg}. In the terminology of~\cite{VonHippel2020}, ``\textsc{There-exists}'' refers to attackers that cannot always lead protocols to a violation of required properties, but \textit{sometimes} succeed (``there exists'' a winning run for the attacker). \cite{VonHippel2020}~formulates the properties that protocols must protect against as \emph{threat models}, and it illustrates its methodology with the Transmission Control Protocol (TCP), specifically connnection establishment using three-way handshake, as standardized in~\cite{rfc793}. The formal model in~\cite{VonHippel2020} was inspired by that in~\cite{jero2015leveraging} where automated attack discovery for TCP is performed using a state-machine-informed search.

In this paper, we revisit the respective ABP and TCP models of~\cite{Alur2017} and~\cite{VonHippel2020} in the standard framework of DES modelled as FSA. In contrast to the feedback-loop control system architecture in the previously-mentioned works on sensor/actuator deception attacks in SCT, we consider a network system architecture in which two peers are sending and receiving packets through channels and/or networks, as explained in~\cref{sec:models}. We consider ``person-in-the-middle'' (PITM) attacks as in~\cite{VonHippel2020,jero2015leveraging}, in a manner reminiscent of deception attacks. Inspired by and complementary to the approach in~\cite{VonHippel2020}, we exploit results in SCT and develop a methodology to synthesize ``\textsc{For-all}'' attackers, that is, attackers that \emph{can always eventually} cause a violation of required properties of the system, extending the previous work by \cite{VonHippel2020} on \textsc{There-exists} attackers.
\cref{sec:procedure} will present the details of our methodology, and will state our main results as \cref{thm:assp}.
We then apply this methodology to both ABP and TCP, using essentially the same models as in~\cite{Alur2017} and~\cite{VonHippel2020}. Thus, our results extend those in~\cite{VonHippel2020} by formally considering the synthesis of ``\textsc{For-all}'' attackers on TCP, since \textsc{For-all} attacks are more powerful than \textsc{There-exists} attacks. In both of our case studies, we approach attack synthesis as a supervisory control problem under partial observation \textit{from the attacker's viewpoint}, which is then solved using existing techniques~\citep{Cassandras2008,wonham_supervisory_2019}. As specifically discussed in \cref{subsec:formulation}, under the assumptions of our PITM attack model, a ``\textsc{For-all}'' attacker for a given threat model is obtained by building the realization of the (partial-observation) \textit{nonblocking} supervisor that results in the supremal controllable and normal sublanguage (supCN) of the \textit{threat model} language with respect to the system language and to the attacker's controllable and observable event sets. The supCN operation was first introduced in~\cite{cho1989supremal}, and several formulas to compute supCN were derived in~\cite{brandt1990formulas}. For each of the two protocols ABP and TCP, respectively in \cref{sec:abp,sec:tcp}, we analyze several setups capturing different PITM attacker capabilities.

The detailed case studies presented in this paper, based upon established models of ABP and TCP (three-way handshake part), show the various steps on how to build, in a systematic manner, successful PITM attacks (if they exist) on these two well-know protocols. We believe they can also serve as inspiration for similar case studies on other protocols.

The remainder of this paper is organized as follows. \cref{sec:preliminaries} provides a brief review of the DES framework and its Supervisory Control Theory employed in this paper. In~\cref{sec:models}, we introduce the context on modelling of communication protocols and give an overview of the PITM attack model under consideration, which is based on specifying a safety or nonblockingness property that the attacker is intent on violating in the context of SCT. \cref{sec:procedure} formulates the SCT-based synthesis problem of a \textsc{For-all} attacker (if it exists) and presents the features of the common methodology that is used in the subsequent sections on ABP and TCP, respectively. ABP is considered first in~\cref{sec:abp}, and then TCP is considered in~\cref{sec:tcp}. Both sections contain sufficient details so that these case studies can be replicated. Finally, we conclude the paper in~\cref{sec:conclusions}.

    \section{Preliminaries}\label{sec:preliminaries}

In this section, we introduce several notions of the DES framework in
\cite{Cassandras2008}, leveraged to build our models in this paper. The central
definitions we need here are \emph{automata}, \emph{nonblockingness} of automata,
\emph{parallel composition}, \emph{supervisory control theory} and
\emph{nonblocking supervisor}.

In DES, what happens in the system is explained by sequences of predefined \emph{events} which discretely occur. Specifically, the system's behaviour is represented as a set of sequences of events, called a \emph{language}, and each sequence is called a \emph{string}. Namely, a language is a set of strings. Note that strings could be arbitrary long and languages could be infinite sets.

One of the intuitive methods to represent (regular) languages is \emph{finite state automata} (FSA), or simply \emph{automata}, represented as a quintuple
\begin{equation}\label{eq:automaton}
    G = (X, E, f, x_0, X_m)
\end{equation}
where $X$ is the finite set of states, $E$ is the finite set of events, $f: X \times E \to X$ is the (partial) transition function, $x_0$ is the initial state and $X_m \subseteq X$ is the set of marked states. The function $f$ denotes the system's behaviour as state transitions defined in the automaton $G$, e.g., $f(x, e) = x'$ represents a transition labelled by event $e \in E$ from state $x \in X$ to state $x' \in X$. From \cref{eq:automaton}, the connection between languages and automata is formally defined as the \emph{generated language} $\lang(G) \coloneqq \{s \in E^* \mid \text{$f(x_0, s)$ is defined}\}$.

From the perspective of system control, it makes sense to consider that several behaviours of the system are acceptable or desired. We call the strings denoting acceptable behaviours \emph{marked strings}, and the language consisting of marked strings is called a \emph{marked language}. To represent the marked language associated with $G$, the marked states in $X_m$ come to play that role. Mathematically, the language marked by $G$ is defined by $\mlang(G) \coloneqq \{s \in \lang(G) \mid f(x_0, s) \in X_m\}$. However, depending on the structure of $G$, it may not be guaranteed that the system $G$ can always eventually reach its marked states. In particular, the existence of deadlock and livelock in $G$ can prevent the marked states from being reached. Such a property in DES is called \emph{nonblockingness}. Specifically, $G$ is said to be blocking if $\prefix{\mlang(G)} \subset \lang(G)$ and nonblocking if $\prefix{\mlang(G)} = \lang(G)$. In other words, if $G$ is blocking, then there exists deadlock or livelock in $G$, that is, there exists a state from which the marked states cannot be reached, and vice versa.

In many cases, the systems we analyze consist of several subcomponents, or one
may want to examine at once the entire behaviour of multiple system models. The
DES framework has an operation of automata called \emph{parallel composition} to
build models of entire systems from subsystem models. For example, the parallel
composition $G'$ of system $G_1$ and system $G_2$ is denoted by $G' = G_1
\parallel G_2$. Roughly speaking, a common event in $G_1$ and $G_2$ can only
occur in $G'$ if both $G_1$ and $G_2$ execute it simultaneously. The private
(unshared) events, on the other hand, can be executed in $G'$ whenever feasible
in either $G_1$ or $G_2$. For the detailed definition and properties of parallel
composition, readers are referred to \cite[pp.~81--87]{Cassandras2008}.

\begin{figure}[htp]
    \centering
    \begin{tikzpicture}[semithick, ->, >=Latex]
    \tikzstyle{plant}=[rectangle, draw, minimum width=7em, minimum height=2.5em]
    \sffamily
    \node[plant] (p) at (0, 0) {Plant $G$};
    \node[plant] (sup) at (0, -4em) {Supervisor $S$};

    \draw (p.east) -| (6em, -2em) |- (sup.east);
    \draw (sup.west) -| (-6em, -2em) |- (p.west);
\end{tikzpicture}
     \caption{The feedback loop of supervisory control}
    \label{fig:models:feedback}
\end{figure}
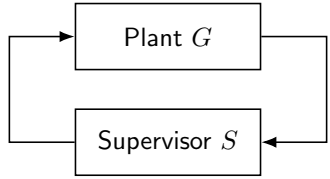

Considering that the given systems do not always follow their specifications,
supervisory control is a concept to control the systems represented as
DES, and its mathematical framework is called \emph{supervisory control theory}
(SCT), which is to synthesize a controller attached to the system so that the
given specifications are satisfied. In the framework of SCT in DES, a system to
be controlled is called a \emph{plant}, and a plant is controlled by a
\emph{supervisor} that enables or disables particular (controllable) events so
that the plant satisfies a given specification for safety or nonblockingness for
instance. The control actions of the supervisor are determined by observation of
the strings generated by the plant; thus the plant and supervisor from a
feedback loop as depicted in \cref{fig:models:feedback}.
Technically speaking, a supervisor $S$ is defined as a function
\begin{equation}\label{eq:supervisor}
    S: \lang(G) \to 2^E
\end{equation}
which takes a string generated by $G$ and returns a set of events permitted to
occur in $G$. In other words, $S(s)$ is a control action for a string $s \in
\lang(G)$. Note that supervisor $S$ is prohibited from disabling a feasible
uncontrollable event at any state. Namely, letting $E_{uc} \subseteq E$ be a
set of uncontrollable events in $G$, for each $s \in \lang(G)$, it always holds
that $E_{uc} \cap \{e \in E \mid \text{$f(f(x_0, s), e)$ is defined}\} \subseteq S(s)$.

In the framework of SCT, it is also considered that the supervisor has a limited observability of events generated by the plant. This limitation is represented by partitioning the set of events $E$ into two disjoint subsets: the sets of observable events $E_o$ and of unobservable events $E_{uo}$, namely $E = E_o \cup E_{uo}$. To implement this property, the supervisor in \cref{eq:supervisor} is extended to the \emph{partial-observation supervisor} $S_P$ defined by
\begin{equation}\label{eq:p-supervisor}
    S_P: P[\lang(G)] \to 2^E
\end{equation}
where $P$ is the natural projection from domain $E^*$ to codomain $E_o^*$, removing unobservable events from a string generated by $G$. Note that in this scheme, the control action by $S_P$ is supposed to always take effect before any unobservable event occurs.

Given $G$ and $S_P$, the closed-loop behaviour of $G$ controlled by $S_P$ is denoted by a DES $S_P/G$, formalized in the following definition.

\begin{defn}[Languages generated and marked by $S_P/G$] (cf.~\cite[p.~151]{Cassandras2008})
    The generated language $\lang(S_P/G)$ is recursively defined as
    \begin{enumerate}
        \item $\varepsilon \in \lang(S_P/G)$
        \item $[s \in \lang(S_P/G) \land s\sigma \in \lang(G) \land \sigma \in S_P[P(s)]] \iff [s\sigma \in \lang(S_P/G)]$
    \end{enumerate}
    and the marked language $\mlang(S_P/G)$ is defined as
    \begin{equation}\label{eq:defn:marklang}
        \mlang(S_P/G) \coloneqq \lang(S_P/G) \cap \mlang(G).
    \end{equation}
    \qed
\end{defn}

We can also examine the blockingness of $S_P/G$ as a meaningful characteristic of the controlled system. Similarly to the blockingness of $G$, the DES $S_P/G$ is said to be \emph{blocking} if $\lang(S_P/G) \neq \prefix{\mlang(S_P/G)}$ and blocking if $\lang(S_P/G) = \prefix{\mlang(S_P/G)}$. Since these properties depend on the synthesis result of $S_P$, $S_P$ is said to be blocking if $S_P/G$ is blocking and to be nonblocking if $S_P/G$ is nonblocking.

The specification that the plant should obey is given as a specification language $L^{spec} \subseteq \lang(G)$, or its automaton representation $H$ such that $\mlang(H) = L^{spec}$. It is an important point that $L^{spec}$ may not be $\mlang(G)$-closed, namely $L^{spec} \neq \prefix{L^{spec}} \cap \mlang(G)$, and we may want the supervisor $S_P$ to ``mark'' strings in $\lang(S_P/G)$ based on $L^{spec}$, rather than $\mlang(G)$. Therefore, the SCT framework provides an alternative version of $S_P$, called a \emph{marking supervisor}, defined as
\begin{equation}\label{eq:marking-sup}
    \mlang(S_P/G) \coloneqq \lang(S_P/G) \cap L^{spec}.
\end{equation}
For the technical details of marking supervisors, the readers are referred to Section 3.9 in \cite{Cassandras2008}. In the rest of this paper, nonblockingness of $S_P$ will be defined by either equation \cref{eq:defn:marklang} or \cref{eq:marking-sup}, depending on the properties of the considered specification $L^{spec}$ (namely, $L^{spec}$ being $\mlang(G)$-closed or not).

    \section{System and Attack Models}\label{sec:models}

Before proceeding to the specific ABP and TCP protocols, we highlight in this section and in the next one the common elements of our two case studies.

\subsection{System Architecture}\label{subsec:models:system}

When modelling communication protocols such as ABP and TCP, we consider a ``system'' that consists of peers communicating with each other, channels, and networks. For clarity of presentation, we suppose the system comprises two peers, two or four channels, and one network. If peers form a small network using channels, e.g., a local area network (LAN), then networks can be omitted and we consider two channels connecting each peer, namely, the forward and backward channels.
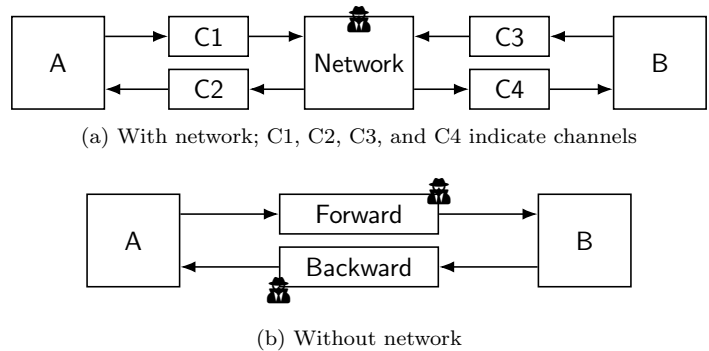
\begin{figure}[htp]
    \centering
    \subcaptionbox{With network; C1, C2, C3, and C4 indicate channels\label{fig:overview:net}}[\linewidth]{
        \begin{tikzpicture}[semithick, ->, >=Latex]
    \sffamily
    \node[peer] (peer_a) at (0, 1em) {A};
    \node[channel] (c1) at (2cm, 2em) {C1};
    \node[channel] (c2) at (2cm, 0) {C2};
    \node[peer] (net) at (4cm, 1em) {Network};
    \node[channel] (c3) at (6cm, 2em) {C3};
    \node[channel] (c4) at (6cm, 0) {C4};
    \node[peer] (peer_b) at (8cm, 1em) {B};
    \node[above=8pt of net.center] (evil) {\smilingimp};

    \draw ($(peer_a.east)+(0,1em)$) -- (c1);
    \draw (c1) -- ($(net.west)+(0,1em)$);
    \draw ($(net.west)-(0,1em)$) -- (c2);
    \draw (c2) -- ($(peer_a.east)-(0,1em)$);
    \draw (c3) -- ($(net.east)+(0,1em)$);
    \draw ($(peer_b.west)+(0,1em)$) -- (c3);
    \draw (c4) -- ($(peer_b.west)-(0,1em)$);
    \draw ($(net.east)-(0,1em)$) -- (c4);
\end{tikzpicture}
     } \\ \medskip
    \subcaptionbox{Without network\label{fig:overview:nonet}}[\linewidth]{
        \begin{tikzpicture}[semithick, ->, >=Latex]
    \sffamily
    \node[peer] (peer_a) at (0, 1em) {A};
    \node[channel, minimum width=6em] (c1) at (3cm, 2em) {Forward};
    \node[channel, minimum width=6em] (c2) at (3cm, 0) {Backward};
    \node[peer] (peer_b) at (6cm, 1em) {B};
    \node[above=0 of c1.east] (evil1) {\smilingimp};
    \node[below=0 of c2.west] (evil2) {\smilingimp};

    \draw ($(peer_a.east)+(0,1em)$) -- (c1);
    \draw (c1) -- ($(peer_b.west)+(0,1em)$);
    \draw ($(peer_b.west)-(0,1em)$) -- (c2);
    \draw (c2) -- ($(peer_a.east)-(0,1em)$);
\end{tikzpicture}
     }
    \caption{Communication overview}
    \label{fig:overview}
\end{figure}
\cref{fig:overview} illustrates an overview of the flow of packets between two peers through channels. Peers A and B exchange packets using communication protocols through the channels and network. In this paper, we consider ``person-in-the-middle'' (PITM) as the attack model on the system. In this model, the attacker \textit{infiltrates the network or channels}, and afterwards sends fake packets and/or discards genuine ones, exploiting vulnerabilities of the protocol (as captured by the peer automata), to damage the system. The system may contain other processes for exogenous events, e.g., timers, called environment processes, which are not depicted in \cref{fig:overview}. Channels work as interfaces between the peers and the network, relaying packets to their destinations. Each component of the system is modelled by a finite-state automaton, and denoted as follows:
\begin{center}
$G_{PA}$: Peer A;
$G_{PB}$: Peer B;
$G_C$: Channel;
$G_N$: Network;
and
$G_e$: Environment processes.
\end{center}
Each channel is represented by one finite-state automaton, thus $G_C$ is the parallel composition of the channel automata. For example, if the system architecture is that in~\cref{fig:overview:net}, then $G_C = G_{C1} \parallel G_{C2} \parallel G_{C3} \parallel G_{C4}$ where $G_{Ci}$ ($i = [1, 4]$) are the respective automata modelling each channel. If the system architecture is that in~\cref{fig:overview:nonet}, then $G_C = G_{FC} \parallel G_{BC}$ where $G_{FC}$ and $G_{BC}$ are the forward and backward channels, respectively, and $G_N$ is empty since there is no network in such an architecture. In the case where there exist more than two environment processes in the system, $G_e$ is also constructed as the parallel composition of all environment processes.

To capture PITM attacks on the above system, we create new versions of the channels and network automata when they are infiltrated by the attacker and denote them by $G_{C,a}$ and $G_{N,a}$, respectively. We consider that the attacker cannot directly tamper the internal codes of peers in our model of PITM attacks, meaning that the attacker cannot disable nor enable the private events of the peers. Instead, in the infiltrated channels or network, the attacker intercepts packets and can delete them, and can also insert new packets to impersonate the sender or receiver, as similarly considered in \cite{jero2015leveraging}. Thus, we construct $G_{C,a}$ and $G_{N,a}$ by the \textit{addition of new transitions and events} that represent the feasible actions of the attacker, as the addition can capture insertion and replacement of packets, and packet deletion by the attacker can be captured by disabling transitions which indicate packet transfer. Concrete examples of $G_{C,a}$ and $G_{N,a}$ will be presented in the case studies in~\cref{sec:abp,sec:tcp}.

Let us define a nominal system model (i.e., without attacker) by
\begin{equation}\label{eq:plant:nominal}
    G_{nom} \coloneqq (X_{nom}, E_{nom}, f_{nom}, x_{nom,0}, X_{nom,m}).
\end{equation}
$G_{nom}$ is the parallel composition of the peers, channels, network, and
environment processes, namely
\begin{equation}\label{eq:plant:nominal:parallel}
    G_{nom} = G_{PA} \parallel G_{PB} \parallel G_C \parallel G_N \parallel G_e
\end{equation}
As we consider PITM attacks on the system, we enhance $G_{nom}$ to the new
model of the system under attack
\begin{equation}\label{eq:plant:attack}
    G_a \coloneqq (X_a, E_a, f_a, x_{a,0}, X_{a,m})
\end{equation}
where possible new transitions and events representing the actions of the attacker
come from the enhanced $G_{C,a}$ and $G_{N,a}$ automata described above.
The other compoents of $G_{nom}$, namely the peer automata $G_{PA}$ and $G_{PB}$, as well as $G_e$, remain unchanged.
In our case studies, the plant $G_a$ is acted upon by the attacker;
hence, the plant consists of the entire system under attack:
\[ G_a = G_{PA} \parallel G_{PB} \parallel G_{C,a} \parallel G_{N,a} \parallel G_e \]

The sending and receiving of packets are represented by events. As we consider PITM attacks, it is reasonable to assume that an attacker infiltrating the network or channels can only monitor incoming and outgoing packets at the infiltrated component. In other words, the attacker cannot observe the private events of the peers. Therefore, we consider that the events in our system model are partitioned into \emph{observable events} and \emph{unobservable events}, based on the system structure and the capability of the attacker. It is also natural to assume that the attacker cannot prevent the peers from sending packets to the network or channels, although the attacker can discard their packets. That is, the attacker cannot control the receiving of packets by the network or channels.

\begin{exmp}\label{exmp:models:abp}
    Let us consider PITM attacks on the Alternating Bit Protocol (ABP). ABP is a protocol which defines the communication mechanism between two peers depicted in \cref{fig:overview:nonet}. Each peer sends and receives packets from its counterpart through the forward and backward channels using first-in-first-out (FIFO) semantics. Inspired by \cite{Alur2017}, we consider $G_{nom}$ as the parallel composition of the following 7 automata.
    \begin{itemize}
        \item $G_S = (X_S, E_S, f_S, x_{S,0}, X_{S,m})$: ABP sender
        \item $G_R = (X_R, E_R, f_R, x_{R,0}, X_{R,m})$: ABP receiver
        \item $G_{FC} = (X_{FC}, E_{FC}, f_{FC}, x_{FC,0}, X_{FC,m})$: Forward channel
        \item $G_{BC} = (X_{BC}, E_{BC}, f_{BC}, x_{BC,0}, X_{BC,m})$: Backward channel
        \item $G_{SC} = (X_{SC}, E_{SC}, f_{SC}, x_{SC,0}, X_{SC,m})$: Sending client
        \item $G_{RC} = (X_{RC}, E_{RC}, f_{RC}, x_{RC,0}, X_{RC,m})$: Receiving client
        \item $G_T = (X_T, E_T, f_T, x_{T,0}, X_{T,m})$: Timer
    \end{itemize}
    Therefore, we have
    \begin{equation}\label{eq:plant:abp:parallel}
        G_{nom} = G_S \parallel G_R \parallel G_{FC} \parallel G_{BC} \parallel
        G_{SC} \parallel G_{RC} \parallel G_T
    \end{equation}
    We also consider that Peer A first sends packets to Peer B, and afterwards Peer B sends an acknowledgement to Peer A. Since Peer A plays a role of the sender side and Peer B is at the receiver side, $G_{PA} = G_S$, $G_{PB} = G_R$, $G_C = G_{FC} \parallel G_{BC}$, and $G_e = G_{SC} \parallel G_{RC} \parallel G_T$, thus \cref{eq:plant:abp:parallel} reduces to \cref{eq:plant:nominal:parallel}. Note that $G_N$ in \cref{eq:plant:nominal:parallel} will be empty in this case.

    The various event sets are defined as follows, where synchronization in $||$ will be achieved by common events:
    \begin{align}
        E_S    & = \{send, done, timeout, p_0, p_1, a_0', a_1'\} \\
        E_R    & = \{deliver, p_0', p_1', a_0, a_1\}             \\
        E_{FC} & = \{p_0, p_1, p_0', p_1'\}                      \\
        E_{BC} & = \{a_0, a_1, a_0', a_1'\}                      \\
        E_{SC} & = \{send, done\}                                \\
        E_{RC} & = \{deliver\}                                   \\
        E_T    & = \{timeout\}
    \end{align}
    Hence
    \begin{align}
        E_{nom} & = E_S \cup E_R \cup E_{FC} \cup E_{BC} \cup E_{SC} \cup E_{RC} \cup E_T         \\
                & = \{send, done, timeout, deliver, p_0, p_1, p_0', p_1', a_0, a_1, a_0', a_1'\}.
    \end{align}
    The events with prefix ``p'' indicate that a packet with indicator bit ``0'' or ``1'' has been sent from the ABP sender to the ABP receiver (i.e., from Peer A to Peer B), and prefix ``a'' indicates an acknowledgement sent from the ABP receiver to the ABP sender, corresponding to which ``0'' or ``1'' has been received by the ABP receiver. The prime symbol is attached to the events of packets and acknowledgement to distinguish those before going through the channel from the corresponding ones after the channels, as is done in~\cite{Alur2017}.

    \cref{fig:abp:models} shows the models of the ABP components. $G_S$ and $G_R$ are example solutions of the distributed protocol completion problem in~\cite{Alur2017}. Note that we have removed from the models in \cref{fig:abp:models} ``dead'' transitions which are never executed by the system when the attacker is not present. The terminology ``dead'' is from \cite{Alur2017}. In addition, we mark all the states of the ABP components, for reasons that will become clear later. Namely,
    \[
        X_{S,m} = X_S,\ X_{R,m} = X_R,\ X_{FC,m} = X_{FC},
        \ X_{BC,m} = X_{BC},\ X_{SC,m} = X_{SC},\ X_{RC,m} = X_{RC},
    \ X_{T,m} = X_T
    \]
    In~\cite{Alur2017}, the forward and backward channels are modelled as nondeterministic finite-state automata as shown in \cref{fig:abp:models:forward,fig:abp:models:backward}. That nondeterminism is introduced to model nonadversarial errors in communication channels, such as packet drop and duplication (see Section 4.2 in~\cite{Alur2017}). To construct the system model in \cref{eq:plant:abp:parallel}, we need deterministic finite-state automata as factors of the parallel composition. Thus, we construct $G_{FC}$ and $G_{BC}$ as observer automata of $G_{FC}^{nd}$ and $G_{BC}^{nd}$, depicted in~\cref{fig:abp:obs-channels}:
\begin{align}
        G_{FC} & = Obs(G_{FC}^{nd}) \label{eq:obs:forward} \\
        G_{BC} & = Obs(G_{BC}^{nd}) \label{eq:obs:backward}
    \end{align}
    where ``observers'' are as defined in \cite{Cassandras2008} and capture the standard conversion of a nondeterministic automaton to a deterministic one (often referred to as subset construction). Observe that $G_{FC}$ and $G_{BC}$ generate exactly the same languages as $G_{FC}^{nd}$ and $G_{BC}^{nd}$, respectively.

    Let us consider one example case of PITM attacks where a powerful attacker infiltrates the forward channel. To construct the plant under attack $G_a$ capturing the attacker's actions, we enhance $G_{FC}^{nd}$ to $G_{FC,a}^{nd}$ as depicted in \cref{fig:abp:mitm:forward} by adding the new transitions shown as the red arrows. This enhanced channel model represents the attacker's capability that can send packets to the recipient with whichever bit 0 or 1, regardless of the incoming packets from the sender. Letting $G_{FC,a} = Obs(G_{FC,a}^{nd})$ in the same way as \cref{eq:obs:forward}, $G_a$ is hereby given by
    \begin{equation}
        G_a = G_S \parallel G_R \parallel G_{FC,a} \parallel G_{BC} \parallel G_e
    \end{equation}
    \cref{sec:abp} describes in detail the procedure to model the PITM attack against ABP.
    \qed
\end{exmp}

In our case studies, $G_a$ is the plant and the attacker plays a role of the supervisor; in this context, the specification represents what damage the attacker wants to cause to the system. In other words, the specification should capture \textit{violations} of a desired property of the communication protocol, such as absence of deadlock or proper delivery of packets. Therefore, using SCT to synthesize a supervisor that enforces the violation of a desired property of the communication protocol under consideration means that we have actually synthesized an attack strategy that indeed causes a violation of that property.

\subsection{\textsc{For-all} Attack}\label{subsec:models:forall}

One of the contributions of this paper as compared to previous work is that we consider that the attacker wants to attack the system in a ``\textsc{For-all}'' manner, to be interpreted in the following sense: \textit{the attacker can always eventually cause a violation of the given property}. Such specifications are naturally captured in SCT using the notion of marked states and nonblockingness. When the marked states capture the violation of the given property, then a \textit{nonblocking supervisory} in SCT will exactly achieve the goal of a \textsc{For-all} attacker, since it will always be possible to eventually reach a marked state. Specifically, consider an attacker's marked (i.e., non-prefix-closed) specification language $L_a^{spec} \subset \lang(G_a)$ which consists of strings that are illegal but feasible in the system under attack. Let $S_a$ be a supervisor (aka attacker) for $G_a$ that achieves as much of $L_a^{spec}$ as possible in the controlled system $S_a/G_a$. We denote this marked language by $K$, namely, $K \subseteq L_a^{spec}$ and the attacker wants $K$ to be as \emph{large} as possible. In order to achieve a \textsc{For-all} attack, the attacker wants $S_a$ to be \textit{nonblocking}, namely, $\mlang(S_a/G_a) = K$ and $\lang(S_a/G_a) = \prefix{K}$. Thus, nonblockingness of the system under attack implies that the attacker can always eventually win; thus, we have indeed obtained a \textsc{For-all} attack strategy. This is how \textsc{For-all} attacks are defined in this paper.

The above definition of \textsc{For-all} attacks is formalized in~\cref{defn:attack}.

\begin{defn}[\textsc{For-all} Attack-Supervisor]\label{defn:attack}
Given $L_a^{spec} \subset \lang(G)$, let $K \subseteq L_a^{spec}$ be a nonempty sublanguage. $S_a$ is said to be a \textsc{For-all} attack-supervisor with respect to $G_a$ and $K$ if
\begin{enumerate}
    \item $\mlang(S_a/G_a) = K$; and
    \item $\lang(S_a/G_a) = \prefix{K}$.
\end{enumerate}
\qed
\end{defn}

\subsection{\textsc{There-exists} Attack}\label{subsec:models:exists}

If there exists a supervisor $S_a$ not satisfying the condition in \cref{defn:attack} but $\lang(S_a/G_a) \cap K \neq \varnothing$, then we say that such an $S_a$ achieves a \textsc{There-exists} attack, because in that case the controlled system (under the actions of the attacker) $S_a/G_a$ will contain deadlocks and/or livelocks (i.e., the system under attack is blocking in the terminology of SCT); this prohibits the attacker from always being able to eventually win. Still, the nonemptyness of $\mlang(S_a/G_a)$ means that the attacker can sometimes win. This is how \textsc{There-exists} attacks are defined in this paper.

The above definition of \textsc{There-exists} attacks is formalized in~\cref{defn:attack:exists}.

\begin{defn}[\textsc{There-exists} Attack-Supervisor]\label{defn:attack:exists}
Given $L_a^{spec} \subset \lang(G)$, let $K \subseteq L_a^{spec}$ be a nonempty sublanguage. $S_a$ is said to be a \textsc{There-exists} attack-supervisor with respect to $G_a$ and $K$ if
\begin{enumerate}
    \item $\lang(S_a/G_a) \cap K \neq \varnothing$; and
    \item $S_a$ is not a \textsc{For-all} attack-supervisor.
\end{enumerate}
\qed
\end{defn}

Now that we have shown how to build the plant model $G_a$, we address in the next section the construction of an automaton representation for the (non-prefix-closed) language $L_a^{spec}$, which will be the ``specification automaton'' for the attacker that is needed in the context of SCT algorithmic procedures.

\begin{rem}
    In the prior work \citep{VonHippel2020}, mostly analogous definitions of \textsc{There-exists} and \textsc{For-all} attackers are given, but in the framework of reactive synthesis with infinite strings and temporal logic (LTL) specifications (see Definition 6 in \cite{VonHippel2020}). The technical difference comes from requiring ``can always eventually win'' instead of requiring ``will always eventually win'' (as is typically done in LTL and is done in \cite{VonHippel2020}). The latter is expressible in LTL, but not the former. The reactive synthesis setting is formally compared to that of SCT in \cite{gap2017jdeds}, where it is shown that nonblockingness in SCT is not expressible in LTL but instead corresponds to ``AGEF(marked)'' in CTL. In this paper, since we use SCT, we match the notion of ``AGEF(marked)'', i.e., ``can always eventually win''. Moreover, since we are working in the context of SCT, we will use the term ``nonblockingness'' for the class of ``liveness'' properties that will be considered in this paper. \qed
\end{rem}

    \section{Procedure for Synthesis of \textsc{For-all} Attacks on Communication Protocols}\label{sec:procedure}

In this section, we discuss the modelling procedure to construct a specification automaton for the attacker based on the considered properties (instances of safety or nonblockingness) of the communication protocol that are to be violated by actions of the attacker. Then, we formulate the problem of finding \textsc{For-all} feasible attacks on the system as a supervisor synthesis problem in SCT which has a known solution. The SCT-based methodology presented in this section will be applied to ABP and TCP in the next two sections.

\subsection{Safety properties}\label{subsec:safety}

As in \cite{Alur2017}, consider a safety property whose violation is modelled by an automaton, termed a \emph{safety monitor} $G_{sm}$. $G_{sm}$ captures the violation of the given safety property in terms of \emph{illegal} states in its structure. Since the specification for attackers represents a violation of the property, the illegal states are represented by \emph{marked} states in $G_{sm}$. In other words, $G_{sm}$ captures the \textit{violation} of the safety property of interest when it reaches its \textit{marked states}. (Note that in our problem context, we do require marked states to capture violation of safety properties.)

$G_{sm}$ can be derived from automata composing $G_{nom}$, namely, the peers, channels, or network, by modifying state marking for instance. One can also independently design $G_{sm}$ as a new automaton that we call a \emph{dedicated automaton} in this paper. Both instances will occur in our case studies. For example, in~\cref{sec:abp}, the safety monitors $G_{sm}$ for ABP are given as dedicated automata in~\cref{fig:abp:safety-monitors}. Let $G_{other}$ be the parallel composition of the automata in $G_{nom}$ which are not used to construct $G_{sm}$. For example, from \cref{eq:plant:nominal:parallel}, if $G_{sm}$ is built by modifying $G_{PA} \parallel G_{PB}$, then $G_{other} = G_C \parallel G_N \parallel G_e$. In~\cref{sec:tcp}, we will construct $G_{sm}$ for the TCP case study using TCP peer models $G_{PA}$ and $G_{PB}$ in~\cref{fig:tcp:peers:timeout} later on.

The specification automaton will in our case studies be the parallel composition of $G_{other}$ and $G_{sm}$, as is commonly done in SCT. Letting $H_{nom}$ be the specification automaton with respect to $G_{nom}$ (system without attacker), we have that $H_{nom} = G_{other} \parallel G_{sm}$. Note that since we want marking in $H_{nom}$ to be determined by marking in $G_{sm}$, all the states of $G_{other}$ are to be marked. In the absence of attackers, the communication protocol should ensure the safety property under consideration, which means that its violation should never occur. This can be verified by confirming that $H_{nom}$ has no reachable marked states, i.e., $H_{nom}$ captures no violations of the given safety property with respect to $G_{nom}$.

To represent the specification automaton with respect to the system under attack, namely $G_a$, we construct $G_{other,a}$ based on $G_a$ in the same manner as $G_{other}$. For instance, if $G_{sm}$ is a dedicated automaton and the attacker infiltrates the network, then $G_{other,a} = G_{PA} \parallel G_{PB} \parallel G_{C} \parallel G_{N,a} \parallel G_e$. Let $H_a = G_{other,a} \parallel G_{sm}$ be the specification automaton under attack. Similarly to marking in $G_{other}$, we want $G_{sm}$ to determine marking in $H_a$, thus all the states of $G_{other,a}$ are to be marked. If there exist no marked states in $H_a$, then the attacker is not powerful enough to cause a violation of the safety property. Even if $H_a$ has marked states, there may not exist \textsc{For-all} attacks (but possibly only \textsc{There-exists} attacks), depending on whether a nonblocking supervisor can be synthesized with respect to plant $G_a$ and specification automaton $H_a$; this will be addressed in the solution of the SCT problem discussed below.

In summary, the procedure to build $H_a$ for a given safety property is presented in \cref{alg:attack-spec:safety}.
\begin{algorithm}[htp]
    \caption{Attack Specification against Safety (\textsc{SafeSpec})}\label{alg:attack-spec:safety}
    \begin{algorithmic}[1]
        \Require{$G_{nom}$, $G_a$, $G_{sm}$}
        \Ensure{$H_a$}
        \If{$G_{sm}$ is a dedicated automaton}
            \State{$G_{other} = G_{nom}$}
            \State{$G_{other,a} = G_a$}
        \Else
            \State{$\Phi = \{G_{PA}, G_{PB}, G_C, G_N, G_e\}$}
            \State{$\Phi_a = \{G_{PA}, G_{PB}, G_{C,a}, G_{N,a}, G_e\}$}
            \State{$G_{other} = \bigparallel \{G \in \Phi \mid \text{$G$ is not used to construct $G_{sm}$}\}$}
            \State{$G_{other,a} = \bigparallel \{G \in \Phi_a \mid \text{$G$ is not used to construct $G_{sm}$}\}$}
        \EndIf
        \State{$H_{nom} = (Y_{nom}, E_{nom}, g_{nom}, y_{nom,0}, Y_{nom,m}) = G_{other} \parallel G_{sm}$}
        \If{$Y_{nom,m} \neq \varnothing$}
            \State{Terminate with empty solution} \Comment{{\small The given model is incorrect as the safety property is violated even if no attacker is present.}}
        \EndIf
        \State{Mark all the states in $G_{other,a}$}
        \State{$H_a = (Y_a, E_a, g_a, y_{a,0}, Y_{a,m}) = Trim(G_{other,a} \parallel G_{sm})$} \Comment{{\small $H_a$ should be trim because we want the attacker to always be able to eventually win, i.e., there should not be any deadlocks/livelocks in the controlled $G_a$.}}
        \If{$Y_{a,m} = \varnothing$}
            \State{Terminate with empty solution} \Comment{{\small The attacker's actions can never cause a violation of the given safety property.}}
        \EndIf
        \State\Return{$H_a$}
    \end{algorithmic}
\end{algorithm}

\begin{prop}\label{prop:attack-spec:safety}
    Suppose that $Y_{nom,m} = \varnothing$ in \cref{alg:attack-spec:safety}, that is, the given system model is correct in terms of the safety properties. If $Y_{a,m}$ on line 16 is empty, then no \textsc{For-all} attack exists and no \textsc{There-exists} attack exists. \qed
\end{prop}

\begin{proof}
    By construction, $G_{sm}$ captures a violation of the given safety property by reaching its marked states. Let $X_{other,a}$ and $X_{sm}$ be the sets of states of $G_{other,a}$ and $G_{sm}$, respectively. Note that $Y_a \subseteq X_{other,a} \times X_{sm}$ from line 15 of \cref{alg:attack-spec:safety}. Since all the states in $X_{other,a}$ are marked, it holds that $Y_{a,m} = \varnothing$ iff for every $(x_{other,a}, x_{sm}) \in Y_a$, $x_{sm}$ is not marked. This means that the safety monitor $G_{sm}$ never captures the violation iff $H_a$ has no marked states. In other words, the attacker can never cause a violation of the given safety property. Therefore, if $Y_{a,m} = \varnothing$, then no \textsc{For-all} attack exists and no \textsc{There-exists} attack exists.
\end{proof}

We build several instances of $H_a$ for ABP in~\cref{subsec:abp:exam} and for TCP in~\cref{subsec:tcp:exam}. The safety monitors for ABP are given as dedicated automata in~\cref{fig:abp:safety-monitors}, as will be explained in~\cref{subsec:abp:safety}, while those for TCP are derived from $G_a$ based on the given safety property, as will be explained in~\cref{subsec:tcp:safety}.

\subsection{Nonblockingness properties}\label{subsec:liveness}

We examine a ``limited'' liveness property, called \emph{nonblockingness}, as expressible in SCT for $*$-languages, namely, languages of finite strings. Nonblockingness is an adequate tool in many applications, such as in software systems; see, e.g.: deadlock in database concurrency control~\citep{lafortune1988modeling}; deadlock in multithreaded programs (Gadara project)~\citep{liao2013eliminating}. Since our approach is based on SCT, nonblockingness is the only type of liveness property that we consider in our case studies. Thus, the set of marked states used for nonblockingness will be the ``parameter'' that captures the desired instance of liveness. In our setting, in \textsc{For-all} attacks the attacker \textit{wants} to cause a violation of nonblockingness with respect to the chosen marked states. First of all, $G_{nom}$ in \cref{eq:plant:nominal:parallel} should be trim for correctness of the system without attacker, as otherwise $G_{nom}$ would contain deadlocks or livelocks. However, $G_a$ should \textit{not} be trim, meaning that the system under attack should contain deadlock or livelock states, i.e., be blocking.

As for the case of safety monitors previously considered, in several instances the violation of the nonblockingness property of interest will be modelled using a dedicated automaton, the \emph{nonblockingness monitor} $G_{nm}$; one such example is shown in~\cref{fig:abp:liveness-monitor}, inspired by \cite{Alur2017} and considered in in~\cref{subsec:abp:liveness}. The \textit{marked} states of $G_{nm}$ will record the \textit{violation} of the given nonblockingness property.

On the other hand, if $G_{nm}$ is not given \textit{a priori}, then violations of nonblockingness will be captured as follows: starting from $G_a$, unmark all states and mark instead the desired (from the viewpoint of the attacker) deadlock and livelock states in $G_a$, resulting in a suitable $G_{nm}$ model. This is done because deadlock and livelock states are illegal, and the attacker wants the system to reach those illegal states (some or all of them, depending on the type of attack). This is the approach that we will follow in our case study on TCP, as will be explained in~\cref{subsubsec:tcp:tm2,subsusbec:tcp:tm3}.

Next, we construct $G_{other,a}$ in the same way as in~\cref{subsec:safety}. That is, we model $G_{other,a}$ as the parallel composition of the automata in $G_a$ which are not used to build $G_{nm}$, and ensure that all the states in $G_{other,a}$ are marked. Note that if $G_{nm}$ is not given as a dedicated automaton and we derive $G_{nm}$ from $G_a$, then $G_{other,a}$ is empty.

Finally, we define $H_a = Trim(G_{other,a} \parallel G_{nm})$, to represent the specification for the attacker which leads the plant to deadlock or livelock states. As a result, we introduce the algorithm to construct $H_a$ in the case of the nonblockingness properties in \cref{alg:attack-spec:liveness}.

\begin{algorithm}[htp]
    \caption{Attack Specification against Nonblockingness (\textsc{NonblockSpec})}\label{alg:attack-spec:liveness}
    \begin{algorithmic}[1]
        \Require{$G_{nom}$, $G_a$, $G_{nm}$}
        \Ensure{$H_a$}
        \If{$G_{nom}$ is not trim}
            \State{Terminate with empty solution} \Comment{{\small The given model is incorrect as the nonblockingness property is violated even if no attacker is present.}}
        \EndIf
        \If{$G_{nm}$ is empty} \Comment{{\small $G_{nm}$ is not given a priori.}}
            \If{$G_a$ is trim}
                \State{Terminate with empty solution} \Comment{{\small The attacker's actions in $G_{C,a}$ and/or $G_{N,a}$ cannot cause a violation of the nonblockingness properties.}}
            \Else
                \State{$X_{a,m} = \varnothing$}
                \State{Add target deadlock/livelock states in $X_a$ to $X_{a,m}$} \Comment{{\small Pick the desired (from the viewpoint of the attacker) deadlock and livelock states.}}
                \State{$G_{nm} = G_a$}
                \State{Let $G_{other,a}$ be empty}
            \EndIf
        \Else
            \State{$\Phi_a = \{G_{PA}, G_{PB}, G_{C,a}, G_{N,a}, G_e\}$}
            \State{$G_{other,a} = \bigparallel \{G \in \Phi_a \mid \text{$G$ is not used to construct $G_{nm}$}\}$}
        \EndIf
        \State{Mark all the states in $G_{other,a}$}
        \State{$H_a = (Y_a, E_a, g_a, y_{a,0}, Y_{a,m}) = Trim(G_{other,a} \parallel G_{nm})$}
        \If{$Y_{a,m} = \varnothing$}
            \State{Terminate with empty solution} \Comment{{\small The attacker's actions can never cause a violation.}}
        \EndIf
        \State\Return{$H_a$}
    \end{algorithmic}
\end{algorithm}

\begin{prop}\label{prop:attack-spec:liveness}
    Suppose that $G_{nom}$ is trim in \cref{alg:attack-spec:liveness}, that is, the given system model is correct in terms of the nonblockingness properties. If $Y_{a,m}$ on line 19 is empty, then no \textsc{For-all} attack exists and no \textsc{There-exists} attack exists.
    \qed
\end{prop}
\begin{proof}
    The proof can be done in the same manner as of \cref{prop:attack-spec:safety}, replacing $G_{sm}$ by $G_{nm}$.
\end{proof}

We will discuss several instances of $H_a$ for ABP in~\cref{subsec:abp:exam} and TCP in~\cref{subsec:tcp:exam}.

\subsection{Problem formulation}\label{subsec:formulation}

In this section, we formulate the Attack-Supervisor Synthesis Problem (ASSP), which is an instance of a standard SCT partial-observation supervisory control problem, but where the attacker plays the role of ``supervisor'' and the specification is a \textit{violation} of a given communication protocol property. ASSP is the formal statement of the \textsc{For-all} attack synthesis problem that is solved in our case studies on ABP and TCP.

\descpar{Attacked-Plant}
As was described earlier, $G_C$ and/or $G_N$ are modified to represent the attacker's ability of inserting and/or discarding packets, resulting in new automata denoted by $G_{C,a}$ and $G_{N,a}$. Next, we form  the plant $G_a$ for ASSP as the parallel composition of nominal and infiltrated automata. For example, if the network is infiltrated by the attacker, then $G_a = G_{PA} \parallel G_{PB} \parallel G_C \parallel G_{N,a} \parallel G_e$.

\descpar{Attack Specification}
Next, we construct $H_a$ using \cref{alg:attack-spec:safety} or \cref{alg:attack-spec:liveness} based on the given safety or nonblockingness property to be violated, as discussed in \cref{subsec:safety} and \cref{subsec:liveness}. Since marking of states in $H_a$ is determined by marking in $G_{sm}$ or $G_{nm}$, the language marked by $H_a$, $\mlang(H_a)$, represents strings where the attacker wins, because
\begin{enumerate}[(i)]
    \item These strings are feasible in $G_a$ by construction.
    \item These strings lead the safety or nonblockingness monitor to a marked state.
\end{enumerate}

As we discussed in \cref{subsec:models:system}, it is reasonable to assume that in PITM attacks the attacker cannot disable or enable the events in the nominal (non-infiltrated) automata, and also that the attacker only observes the events in the automata of the infiltrated components. Thus we define the two partitions of $E_a$ in \cref{eq:plant:attack}, from the viewpoint of the attacker (which plays the role of supervisor):
\begin{enumerate}[(i)]
    \item \emph{Controllable events} $E_{a,c}$ and \emph{uncontrollable events}
          $E_{a,uc}$ for controllability.
    \item \emph{Obsevable events} $E_{a,o}$ and \emph{unobservable events}
          $E_{a,uo}$ for observability.
\end{enumerate}

Consequently, we have the following supervisory control problem, under partial observation, for the attacker.

\begin{prob}[Attack-Supervisor Synthesis Problem, or ASSP]\label{prob:assp}
    Let $G_a$ be a plant automaton, under attack, as in~\cref{eq:plant:attack}; $E_{a,c}$ be a set of controllable events; $E_{a,o}$ be a set of observable events; and $\mlang(H_a) \subset \lang(G_a)$ be a marked (non-prefix-closed) specification language. Find a \textit{maximal controllable and observable} sublanguage of $\mlang(H_a)$ with respect to $\lang(G_a)$, $E_{a,c}$, and $E_{a,o}$, if a non-empty one exists.
    \qed
\end{prob}

The following theorem states that a non-empty output of ASSP will be the controlled behaviour under a successful \textsc{For-all} attack, highlighting our main results in this paper.

\begin{thm}\label{thm:assp}
Let $K$ be a solution of ASSP. Then there exists a \textsc{For-all} attack-supervisor with respect to $G_a$ and $K$. Conversely, if ASSP has no non-empty solution, then there does not exist a \textsc{For-all} attack-supervisor for $L^{spec}_a = \mlang(H_a)$, with the given controllable and observable event sets for the attacker.
\qed
\end{thm}
\begin{proof}
Since $K$ is a controllable and observable sublanguage of $\mlang(H_a) \subset \lang(G_a)$, from the ``controllability and observability theorem'' \citep[p.~197]{Cassandras2008}, there exists a supervisor $S_P$ such that $\mlang(S_P/G_a) = K$ and $\lang(S_P/G_a) = \prefix{K}$. From \cref{defn:attack}, $S_P$ here is a \textsc{For-all} attack-supervisor with respect to $G_a$ and $K$. If $\mlang(H_a)$ is not $\mlang(G)$-closed, we consider $S_P$ to be a marking supervisor, as mentioned in \cref{sec:preliminaries}. Conversely, if the empty set is the only solution to ASSP, then there is no \textsc{For-all} attacker: this is because there is no non-empty language satisfying conditions 1 and 2 in \cref{defn:attack}.
\end{proof}

The \textit{realization} (using standard SCT terminology) of the corresponding (nonblocking) supervisor will encode the control actions of the attacker. By taking the parallel composition of the supervisor's realization with the plant, we obtain an automaton that is language equivalent (generated and marked) to the plant under supervision. Namely, letting $R_a$ be the realization of $S_P$, it holds that $R_a \parallel G_a$ is language equivalent to the controlled plant $S_P/G_a$; see \cite{Cassandras2008,wonham_supervisory_2019}. $R_a$ therefore corresponds to a TM-attacker as defined in~\cite{VonHippel2020}. In ASSP, we require maximalty of the controllable and observable sublanguage, since this problem is known to be solvable \citep{yin2016tac}.

In the PITM attack model, the assumption of $E_{a,c} \subseteq E_{a,o}$ usually holds. In fact, in all of the scenarios considered in~\cref{sec:abp,sec:tcp}, the condition $E_{a,c} \subseteq E_{a,o}$ will hold. In this important special case, the supremal controllable and observable sublanguage of $\mlang(H_a)$ with respect to $\lang(G_a)$, $E_{a,c}$, and $E_{a,o}$ exists and is equal to the supremal controllable and normal sublanguage of $\mlang(H_a)$, denoted by $\supcn{\mlang(H_a)}$, with respect to $\lang(G_a)$, $E_{a,c}$, and $E_{a,o}$. If it is empty, then no \textsc{For-all} attack exists for the given safety or nonblockingness property.

If $\supcn{\mlang(H_a)} \neq \varnothing$, then this language represents the largest attacked behaviour which is possible in the context of a \textsc{For-all} attack against the safety or nonblockingness property. Any marked string in that language provides an example of a successful attack, which is feasible in $G_a$ and steers $G_{nm}$ or $G_{sm}$ to its marked (illegal) state. Let $H_a^{CN}$ be the trim automaton output by the algorithm for the supremal controllable and normal sublanguage, namely
\begin{equation}
    \mlang(H_a^{CN}) = \supcn{\mlang(H_a)}
\end{equation}
and
\begin{equation}
    \lang(H_a^{CN}) = \prefix{\supcn{\mlang(H_a)}}
\end{equation}
From the controllability and observability theorem of SCT, there exists a partial-observation nonblocking supervisor $S_P$ such that
\begin{equation}\label{eq:p-sup}
    \lang(S_P/G_a) = \prefix{\supcn{\mlang(H_a)}} = \lang(H_a^{CN})
\end{equation}
$S_P$ corresponds to a \textsc{For-all} attack-supervisor since every string in the controlled behaviour, $S_P/G_a$, can be extended to a marked string, by nonblockingness of $S_P$. In other words, it is always eventually possible for the system under attack by $S_P$ to violate the given property.

In the above formulation, $\mlang(H_a)$ may not be $\mlang(G_a)$-closed, since it is possible that $G_a = G_{other,a}$ and all the states in $G_a$ are marked. Therefore, according to the use of $G_{sm}$ and $G_{nm}$, whenever necessary we define $S_P$ as a marking supervisor by following \cref{eq:marking-sup}, namely
\begin{equation}\label{eq:marking-supervisor}
    \mlang(S_P/G_a) \coloneqq \lang(S_P/G_a) \cap \mlang(H_a^{CN}) = \supcn{\mlang(H_a)}
\end{equation}

As a last step, we need to build a realization of $S_P$ as an automaton that: (i) only changes its state upon the occurrence of observable events, since $H_a^{CN}$ contains transitions with unobservable events; and (ii) whose active event set at each state of the realization is equal to the events \textit{enabled} by the supervisor (attacker) at that state. Noting that marking of states may be relevant in the case of a marking supervisor, the standard process for automaton realization of a partial-observation supervisor (see Section 3.7.2 in~\cite{Cassandras2008}) can be followed. From \cref{eq:p-sup,eq:marking-supervisor}, we build an automaton realization of $S_P$ using $H_a^{CN}$, where $S_P$ is such that
\begin{equation}
    \mlang(S_P/G_a) = \supcn{\mlang(H_a)}
\end{equation}
and
\begin{equation}
    \lang(S_P/G_a) = \prefix{\supcn{\mlang(H_a)}}
\end{equation}
First, we build the observer of $H_a^{CN}$, $Obs(H_a^{CN})$, with respect to $E_{a,o}$, using the standard process of observer construction~\citep{Cassandras2008}. Next, we add self loops for all events in $E_{a,c} \cap E_{a,uo}$ that need to be enabled at each state of $Obs(H_a^{CN})$, obtained by examining the corresponding states of $H_a^{CN}$. The attack strategy of the successful \textsc{For-all} attacker is encoded in this realization, as desired.

Based on the above discussion, we introduce \cref{alg:synthesis} to synthesize \textsc{For-all} attacks with respect to the given $G_{nom}$, $G_a$ and $G_m$ (either a safety or nonblockingness monitor). We also state in \cref{prop:synthesis} that \cref{alg:synthesis} returns the realization of a \textsc{For-all} attack-supervisor, if it exists, which encodes the attack strategy in order for the attacker to lead the plant to a violation of the given safety/nonblockingness monitor.

\begin{algorithm}[htp]
    \caption{\textsc{For-all} Attack Synthesis}\label{alg:synthesis}
    \begin{algorithmic}[1]
        \Require $G_{nom}$, $G_a$, $G_m$
        \Ensure $R$
        \If{$G_m$ is a safety monitor}
            \State $H_a = \Call{SafeSpec}{G_{nom},G_a,G_m}$
        \Else
            \State $H_a = \Call{NonblockSpec}{G_{nom},G_a,G_m}$
        \EndIf
        \State Compute $\supcn{\mlang(H_a)} = \mlang(H_a^{CN})$ from $G_a$ and $H_a$ \Comment{{\small $H_a^{CN}$ is the trim automaton output by the standard algorithm \citep{Cassandras2008} for the supremal controllable and normal sublanguage.}}
        \If{$\supcn{\mlang(H_a)}$ is empty}
            \State Terminate with empty solution
        \EndIf
        \State Compute the realization $R$ of $S_P$ from $H_a^{CN}$ such that $\mlang(S_P/G_a) = \supcn{\mlang(H_a)}$ and $\lang(S_P/G_a) = \prefix{\supcn{\mlang(H_a)}}$
        \State \Return $R$
    \end{algorithmic}
\end{algorithm}

\begin{prop}\label{prop:synthesis}
    Suppose that $H_a$ on line 2 or line 4 in \cref{alg:synthesis} is non-empty, i.e., \cref{alg:attack-spec:safety} or \cref{alg:attack-spec:liveness} returns a non-empty solution. If ASSP (\cref{prob:assp}) is solvable, then \cref{alg:synthesis} returns the realization of a \textsc{For-all} attack-supervisor.
    \qed
\end{prop}

\begin{proof}
    Since $E_{a,c} \subseteq E_{a,o}$, if there exists a solution of ASSP, then the supremal controllable and observable sublanguage of $\mlang(H_a)$ exists and is equal to $\supcn{\mlang(H_a)}$, which is a solution of ASSP. Thus from the proof of \cref{thm:assp}, a supervisor $S_P$ such that $\mlang(S_P/G_a) = \supcn{\mlang(H_a)}$ and $\lang(S_P/G_a) = \prefix{\supcn{\mlang(H_a)}}$ is a \textsc{For-all} attack-supervisor. Therefore, if ASSP is solvable, then \cref{alg:synthesis} returns the realization of a \textsc{For-all} attack-supervisor.
\end{proof}

As long as \cref{alg:synthesis} returns a non-empty automaton, from \cref{prop:synthesis}, the above methodology results in a closed-loop system that produces \textsc{For-all} attacks, in the presence of the attacker. Since $H_a^{CN}$ in \cref{alg:synthesis} is a trim automaton, we know that at any state in $H_a^{CN}$, it is possible to reach a marked state, resulting in a violation of the monitor. Therefore, it is always possible for the attacker to eventually win.

\begin{rem}
    When $H_a$ output by \cref{alg:attack-spec:safety} or \cref{alg:attack-spec:liveness} is not empty (i.e., when it has at least one marked state) but there is no \textsc{For-all} attack-supervisor (i.e., \cref{alg:synthesis} returns the empty solution), then we can conclude that there exists at least one \textsc{There-exists} attack-supervisor, according to \cref{defn:attack:exists}. For instance, one can take the supervisor $S_{all}$ that always enables all events. Then $\lang(S_{all}/G_a) = \lang(G_a)$ and $\lang(G_a) \cap \mlang(H_a) = \mlang(H_a)$ by construction of $H_a$. Hence, this attack-supervisor can reach any of the marked states in $H_a$ where it ``wins'', but the closed-loop system will be blocking. Techniques in SCT for synthesizing blocking supervisors, as described in Section 3.5.5 of \cite{Cassandras2008} for instance, can be employed to guide the design of \textsc{There-exists} attack-supervisors when no \textsc{For-all} attack-supervisor exists. Further investigation of \textsc{There-exists} attack-supervisors is beyond the scope of this paper.
\end{rem}

    \section{ABP Case Study}\label{sec:abp}

Our first case study for synthesis of \textsc{For-all} attacks is for the Alternating Bit Protocol (ABP), as studied and modelled in~\cite{Alur2017}. The models of ABP components we use in this section are described in \cref{exmp:models:abp}.

\subsection{Safety property models}\label{subsec:abp:safety}

As introduced in \cref{subsec:safety}, safety properties are represented by
safety monitor automata which define what states in the system must not be
reached, i.e., define illegal states.
\cite{Alur2017}~provides two safety monitor automata, $G_{sm}^1$ and $G_{sm}^2$,
capturing the violation of safety properties for ABP, depicted
in~\cref{fig:abp:safety-monitors}. The marked state $q_2$ in $G_{sm}^1$ and
$G_{sm}^2$ indicates the illegal state, namely, the safety property is violated
if the monitor reaches this state from the initial state. $G_{sm}^1$ expresses that:
\begin{itemize}
    \item $deliver$ should happen after $send$, meaning that $deliver$ of the ABP
          receiver and the Receiving client should not happen before the Sending
          client tells the ABP sender to send a bit to the forward channel.
    \item After $send$ happens, the next $send$ should not occur before $deliver$
          occurs, meaning that the Sending client should wait for the
          acknowledgement signal from the ABP receiver.
\end{itemize}
On the other hand, $G_{sm}^2$ expresses that:
\begin{itemize}
    \item $done$ should happen after $deliver$, meaning that $done$ of the ABP
          sender and the Sending client should not happen before the ABP receiver
          receives the signal and sends the acknowledgement to the ABP sender.
    \item After $deliver$ happens, the next $deliver$ should not occur before
          $done$ occurs, meaning that $deliver$ cannot happen before the Sending
          client tells the ABP sender to send the next signal to the forward
          channel.
\end{itemize}

Since the safety monitors are provided as dedicated automata, $G_{sm}^1$ and $G_{sm}^2$, $G_{other}$ in \cref{alg:attack-spec:safety} is equal to $G_{nom}$. In our ABP system model, $H_{nom}$ on line 10 in \cref{alg:attack-spec:safety} has no marked states, thus we state that our ABP model is correct in terms of the safety properties. Namely, the nominal system (without attacker) does not violate the given safety properties.

\subsection{Nonblockingness property models}\label{subsec:abp:liveness}

The nonblockingness monitor in~\cref{fig:abp:liveness-monitor}, $G_{nm}$,
captures a violation of the nonblockingness
property that the entire system should not get stuck, and should not keep
invoking $send$. Namely, the first $send$ should eventually be followed by a $deliver$.
$G_{nm}$ in~\cref{fig:abp:liveness-monitor} is a simplified version of a monitor provided by \cite{Alur2017} so that our nonblockingness monitor $G_{nm}$ captures that the first transmission is never completed, which is adequate for our case study.

\subsection{Attack model}\label{subsec:abp:attack-model}

As we consider the system architecture in~\cref{fig:overview:nonet} for ABP, the attacker infiltrates the forward and/or backward channels. To follow \cref{alg:attack-spec:safety,alg:attack-spec:liveness}, we first construct a modified model of the plant $G_a$ in~\cref{eq:plant:attack} under attack. Since the channels of ABP are under attack, we enhance $G_{FC}$ and $G_{BC}$ to those under attack, $G_{FC,a}$ and $G_{BC,a}$, by adding new transitions to represent capabilities of the attacker. Note that if we keep either of the channels nominal, then $G_{FC,a} = G_{FC}$ or $G_{BC,a} = G_{BC}$ accordingly. Therefore, $G_{C,a} = G_{FC,a} \parallel G_{BC,a}$.

The PITM attacker is represented by a modified forward or backward channel that can send the recipient a different packet from the incoming packet.  For example, if the attacker has infiltrated the forward channel, then the attacker can send either $p_0'$ or $p_1'$ to the ABP receiver regardless of which $p_0$ or $p_1$ occurs. \cref{fig:abp:mitm} shows the attacked forward and backward channels. Red transitions are added to the original channel models in~\cref{fig:abp:models:forward,fig:abp:models:backward}. These new transitions enable the attacker to send whichever packet they want. To construct $G_a$, we model $G_{FC,a}$ and $G_{BC,a}$ as observer automata of $G_{FC,a}^{nd}$ and $G_{BC,a}^{nd}$, as was done for $G_{nom}$. \cref{fig:abp:mitm:obs} depicts $G_{FC,a}$ and $G_{BC,a}$, representing new transitions compared to~\cref{fig:abp:obs-channels} as red transitions.

As discussed in~\cref{subsec:models:system}, we suppose that the attacker cannot control and observe events outside the channels. Therefore, the event set $E_a$ is partitioned as follows:
\begin{itemize}
    \item Controllable events: $E_{a,c} = \{p_0', p_1', a_0', a_1'\}$
    \item Uncontrollable events: $E_{a,uc} = \{send, done, timeout, deliver, p_0, p_1, a_0, a_1\}$
    \item Observable events: $E_{a,o} = \{p_0, p_1, p_0', p_1', a_0, a_1, a_0', a_1'\}$
    \item Unobservable events: $E_{a,uo} = \{send, done, timeout, deliver\}$.
\end{itemize}
We consider that in our attack model, the attacker controls the output packets from the channels so that each safety or nonblockingness monitor in~\cref{subsec:abp:safety,subsec:abp:liveness} reaches its marked state, if possible.

\subsection{Examination of the PITM attack for ABP}\label{subsec:abp:exam}

In this section, we examine the PITM attack for the above safety and nonblockingness properties of ABP according to the following steps:
\begin{enumerate}
    \item Construct the plant under attack $G_a$ as the parallel composition of the component models of ABP under attack, namely
    \begin{equation}
        G_a = G_S \parallel G_R \parallel G_{C,a} \parallel G_e
    \end{equation}
    where $G_{C,a} = G_{FC,a} \parallel G_{BC,a}$ and $G_e = G_{SC} \parallel G_{RC} \parallel G_T$.
    \item Using \cref{alg:synthesis}, compute the realization of a \textsc{For-all} attack-supervisor with respect to $G_{nom}$, $G_a$ and the safety/nonblockingness monitor for ABP.
\end{enumerate}
For illustration purposes, if \cref{alg:synthesis} returns the realization of an attack-supervisor, we pick one example string from the initial state to one marked state in $\supcn{\mlang(H_a)}$, which represents one system behaviour under attack that reaches a marked state in the monitor.

$G_a$ varies depending on $G_{C,a}$, namely which channel is under the PITM attack, so we consider the following three cases in each setup:
\begin{enumerate}
    \item The forward channel is under the PITM attack (i.e. $G_{BC,a} = G_{BC}$):
          \begin{equation}\label{eq:system:case1}
              G_a = G_S \parallel G_R \parallel G_{FC,a} \parallel G_{BC} \parallel G_e
          \end{equation}
    \item The backward channel is under the PITM attack (i.e. $G_{FC,a} = G_{FC}$):
          \begin{equation}\label{eq:system:case2}
              G_a = G_S \parallel G_R \parallel G_{FC} \parallel G_{BC,a} \parallel G_e
          \end{equation}
    \item Both channels are under the PITM attack:
          \begin{equation}\label{eq:system:case3}
              G_a = G_S \parallel G_R \parallel G_{FC,a} \parallel G_{BC,a} \parallel G_e
          \end{equation}
\end{enumerate}
For clarity of presentation, we henceforth focus on the use of the safety monitor 1 and $G_a$ in \cref{eq:system:case1} in which the forward channel is under attack, as presented in \cref{exmp:models:abp}. In other words, we consider $H_a$ as the parallel composition of $G_a$ in \cref{eq:system:case1} and the safety monitor 1 $G_{sm}^1$. The other cases of \cref{eq:system:case2,eq:system:case3} and the safety monitor~2 can be examined using the same procedure.

\subsubsection{Attack against Safety Properties}\label{subsubsec:abp:safety}

\paragraph{Setup 1}

Consider the PITM channels in~\cref{fig:abp:mitm} which represent a powerful attacker that can send packets to the recipient with whichever bit 0 or 1, regardless of the incoming packets.

Following our procedure, we found that $H_a$ has 168 marked states out of 265 states and $H_a^{CN}$ is non-empty. Here, $\mlang(H_a^{CN}) = \mlang(H_a)$ and $\lang(H_a^{CN}) = \lang(G_a)$, so $H_a$ is already controllable and normal with respect to $G_a$, thus the attacker issues no disablement actions. Let us pick the example string $send.p_0.p_0'.deliver.a_0.p_1'.deliver$, which means that the attacker sends the correct packet with bit 0 first, and afterwards sends a fake packet with bit 1 to the ABP receiver when it observes $a_0$. In other words, the attacker inserts $q_1'$ soon after it observes $a_0$. Consequently, $G_{sm}^1$ captures the violation by reaching $q_2$ with $send.deliver.deliver$.

\paragraph{Setup 2}

Let us represent a less-powerful attacker by removing additional transitions from the PITM channels in~\cref{fig:abp:mitm}. First, we remove all red transitions except $p_1'$ from $f_1$ to $f_0$ in~\cref{fig:abp:mitm:forward}, so that the attacker can send packets with bit 1 at the particular timing. Let $G_{FC,wa}^{nd}$ be the less powerful forward PITM channel derived from $G_{FC,a}^{nd}$. \cref{fig:abp:weak} shows $G_{FC,wa}^{nd}$ and $G_{FC,wa} = Obs(G_{FC,wa}^{nd})$. The red transitions are new ones compared to $G_{FC}^{nd}$ and $G_{FC}$.

Next, we compute $G_a$, $H_a$, and $H_a^{CN}$ by following the steps at the beginning of~\cref{subsec:abp:exam}. $G_a = G_S' \parallel G_R \parallel G_{FC,wa} \parallel G_{BC} \parallel G_e'$ has 248 states, and $H_a = G_a \parallel G_{sm}^1$ has 370 states and 228 marked states. $H_a^{CN}$ is non-empty and consists of 1099 states and 771 marked states. In every case, $\lang(H_a^{CN}) = \lang(G_a)$, so no disabling happens. As the example string in $H_a^{CN}$, we pick $send.p_0.p_0'.deliver.a_0.p_1'.deliver$ which is the same as that in Setup 1, but $H_a^{CN}$ here is not equivalent. Let $(H_a^{CN})_2$ be $H_a^{CN}$ here and $(H_a^{CN})_1$ be $H_a^{CN}$ in Setup 1. Since $(H_a^{CN})_2^{comp} \times (H_a^{CN})_1$ is non-empty, we conclude that $(H_a^{CN})_2$ lacks some attack strategies, but one additional $p_1'$ in $G_{FC,wa}^{nd}$ is enough to cause the violation of the safety property.

\paragraph{Setup 3}

Let us make the attacker much less powerful than in Setup 2, by building a new automaton of the infiltrated forward channel and changing the sets of controllable and observable events.

Consider the new automaton of the infiltrated forward channel, depicted in~\cref{fig:abp:mitm:oneshot}. We denote this new automaton by $G_{FC,a}^{oneshot,nd}$ and its observer by $G_{FC,a}^{oneshot}$, namely $G_{FC,a}^{oneshot} = Obs(G_{FC,a}^{oneshot,nd})$. This forward channel means that the attacker can send a fake packet with bit 1 to the ABP receiver only once (one-shot attacker). After the fake packet, the channel's behaviour will get back to normal. Moreover, we consider the following controllable and observable event sets:
\begin{itemize}
    \item Controllable events: $E_{a,c} = \{p_1'\}$
    \item Observable events: $E_{a,o} = \{p_0, p_1, p_0', p_1', a_0, a_1, a_0', a_1'\}$
\end{itemize}
meaning that the attacker can observe events in both of the channels, but can only control $p_1'$ in the (infiltrated) forward channel. By following the procedure as we have done, $G_a$ in~\cref{eq:system:case1}, where $G_{FC,a} = G_{FC,a}^{oneshot}$, has 334 states. Also, $H_a = G_a \parallel G_{sm1}$ has 190 marked states out of 431 states, and $H_a^{CN}$ is non-empty. Moreover, $\mlang(H_a^{CN}) \neq \mlang(H_a)$ and $\lang(H_a^{CN}) \neq \lang(G_a)$, thus the attacker issues event disablement actions during its attack on the system. For illustration, we pick the following example string in $H_a^{CN}$:
\[
    send.p_0.p_0'.deliver.a_0.a_0'.done.send.p_1.p_1'.deliver.a_1.a_1'.done
    .send.p_0.p_0'.deliver.a_0.{\color{blue}a_1'}.{\color{red}p_1'}.deliver.a_1
\]
By observation, the blue events are nonadversarial error packets which are sent mistakenly, and the red event ${\color{red}p_1'}$ is inserted by the attacker. Note that the attacker can observe $p_1'$ and $a_1'$ here. Accordingly, this string means that the attacker can lead the system to the undesired state by sending the fake packet $p_1'$ only once after the observation of one error packet. Moreover, the attacker disables $p_1'$ several times before sending the fake $p_1'$. Therefore, in this case, the violation is caused ``by chance'', since the attacker exploits errors, but that violation is enabled by the attacker's intervention. It is worth mentioning that if we remove the events in the backward channel (i.e., $a_0$, $a_1$, $a_0'$ and $a_1'$) from $E_{a,o}$, then $H_a^{CN}$ is empty. This means that the attacker needs to observe the behaviour of the backward channel so as to exploit nonadversarial errors to attack. Moreover, if we set $E_{a,c} = \varnothing$ and $E_{a,o} = \{p_0, p_1, p_0', p_1', a_0, a_1, a_0', a_1'\}$, then $H_a^{CN}$ is empty again, meaning that the attacker needs to have the controllability of $p_1'$ to attack successfully.

\subsubsection{Attack against Nonblockingness Properties}\label{subsubsec:abp:liveness}

\paragraph{Setup 4}

Consider that the attacker wants the system to violate the nonblockingness property represented by the nonblockingness monitor $G_{nm}$ in~\cref{fig:abp:liveness-monitor}. Let us examine the system under attack where the forward channels are infiltrated by the attacker, namely $G_a$ in~\cref{eq:system:case1}. Note that the forward PITM channel here is that in~\cref{fig:abp:mitm:forward} which is quite powerful. Since $G_{nm}$ is given as a dedicated automaton, we build $H_a = Trim(G_{other,a} \parallel G_{nm})$ where $G_{other,a} = G_a$.

In this case, $G_a$ consists of 174 states, and $H_a$ comprises 14 states and 13 marked states. $H_a^{CN}$ is non-empty and consists of 10 states and 9 marked states. As the example string in $H_a^{CN}$, we pick string $send.p0.p1'.a1.timeout$ which means that the attacker sends a fake packet with bit 1 to the ABP receiver after it observes $p0$, and expects the system to suffer from timeout. Moreover, from $H_a^{CN}$, the attacker-supervisor disables $p_0'$ to prevent $deliver$, resulting in $\lang(H_a^{CN}) \neq \lang(G_a)$. Therefore, there exist no $deliver$ transitions in $H_a^{CN}$. This result shows that the attacker successfully leads the system to violate the nonblockingness property that $send$ should eventually be followed by $deliver$.

    \section{TCP Case Study}\label{sec:tcp}

Our second case study concerns one of the major protocols in the Internet, the Transmission Control Protocol (TCP)~\citep{rfc793}. TCP is widely used to communicate through unreliable paths. We consider a communication architecture as in~\cref{fig:overview:net}. Each peer sends and receives packets to and from channels, and the network interconnects channels to relay the incoming packets to their destinations. As in~\cite{VonHippel2020}, we consider the connection establishment phase of TCP, based on three-way handshake, and do not model the congestion control part of that protocol.

\subsection{Component models of TCP}\label{subsec:tcp:components}

Let $G_{nom}$ in~\cref{eq:plant:nominal} be the entire connection establishment part of TCP without an attacker. Based on the architecture of TCP introduced in~\cite{VonHippel2020}, we consider $G_{nom}$ as the parallel composition of the following components:
\begin{itemize}
    \item $G_{PA} = (X_{PA}, E_{PA}, f_{PA}, x_{PA,0}, X_{PA,m})$: Peer A
    \item $G_{PB} = (X_{PB}, E_{PB}, f_{PB}, x_{PB,0}, X_{PB,m})$: Peer B
    \item $G_{C1} = (X_{C1}, E_{C1}, f_{C1}, x_{C1,0}, X_{C1,m})$: Channel 1
    \item $G_{C2} = (X_{C2}, E_{C2}, f_{C2}, x_{C2,0}, X_{C2,m})$: Channel 2
    \item $G_{C3} = (X_{C3}, E_{C3}, f_{C3}, x_{C3,0}, X_{C3,m})$: Channel 3
    \item $G_{C4} = (X_{C4}, E_{C4}, f_{C4}, x_{C4,0}, X_{C4,m})$: Channel 4
    \item $G_N = (X_N, E_N, f_N, x_{N,0}, X_{N,m})$: Network
\end{itemize}
namely
\begin{equation}\label{eq:tcp:system:nominal}
    G_{nom} = G_{PA} \parallel G_{PB} \parallel G_{C1} \parallel G_{C2} \parallel G_{C3} \parallel G_{C4} \parallel G_N
\end{equation}
Hence, $G_C = G_{C1} \parallel G_{C2} \parallel G_{C3} \parallel G_{C4}$ and $G_e = G_N$, so~\cref{eq:tcp:system:nominal} reduces to~\cref{eq:plant:nominal:parallel}.

The event sets are defined as follows:
\begin{align}
    E_{PA} &=
    \begin{multlined}[t]
        \{listen_A, timeout_A, deleteTCB_A, SYN_{AC1}, SYN_{C2A}, \\
        ACK_{AC1}, ACK_{C2A}, FIN_{AC1}, FIN_{C2A}, SYN\_ACK_{AC1},
        SYN\_ACK_{C2A}\}
    \end{multlined} \label{eq:tcp:events:peerA}\\
    E_{PB} &=
    \begin{multlined}[t]
        \{listen_B, timeout_B, deleteTCB_B, SYN_{BC3}, SYN_{C4B}, \\
        ACK_{BC3}, ACK_{C4B}, FIN_{BC3}, FIN_{C4B}, SYN\_ACK_{BC3},
        SYN\_ACK_{C4B}\}
    \end{multlined} \label{eq:tcp:events:peerB}\\
    E_{C1} & = \{SYN_{AC1}, SYN_{C1N}, ACK_{AC1}, ACK_{C1N}, FIN_{AC1}, FIN_{C1N}, SYN\_ACK_{AC1}, SYN\_ACK_{C1N}\}              \\
    E_{C2} & = \{SYN_{NC2}, SYN_{C2A}, ACK_{NC2}, ACK_{C2A}, FIN_{NC2}, FIN_{C2A}, SYN\_ACK_{NC2}, SYN\_ACK_{C2A}\}              \\
    E_{C3} & = \{SYN_{BC3}, SYN_{C3N}, ACK_{BC3}, ACK_{C3N}, FIN_{BC3}, FIN_{C3N}, SYN\_ACK_{BC3}, SYN\_ACK_{C3N}\}              \\
    E_{C4} & = \{SYN_{NC4}, SYN_{C4B}, ACK_{NC4}, ACK_{C4B}, FIN_{NC4}, FIN_{C4B}, SYN\_ACK_{NC4}, SYN\_ACK_{C4B}\}              \\
    E_N &=
    \begin{multlined}[t]
        \{SYN_{C1N}, SYN_{C3N}, SYN_{NC2}, SYN_{NC4}, \\
        ACK_{C1N}, ACK_{C3N}, ACK_{NC2}, ACK_{NC4}, \\
        FIN_{C1N}, FIN_{C3N}, FIN_{NC2}, FIN_{NC4}, \\
        SYN\_ACK_{C1N}, SYN\_ACK_{C3N}, SYN\_ACK_{NC2}, SYN\_ACK_{NC4}\}
    \end{multlined} \label{eq:tcp:events:network:nominal}
\end{align}
Hence
\begin{equation}\label{eq:tcp:eventset:nominal}
    E_{nom} = E_{PA} \cup E_{PB} \cup E_{C1} \cup E_{C2} \cup E_{C3} \cup E_{C4} \cup E_N
\end{equation}
The subscripts in the event names indicate the directions of packets. For example, ``AC1'' means packets from Peer A to Channel 1. Note that the subscripts ``A'' and ``B'' are added to ``listen'' and ``deleteTCB'' to make these events private.

\crefrange{fig:tcp:component:2}{fig:tcp:peers:timeout} depict the models of the above TCP components. $G_{PA}$ and $G_{PB}$ illustrate the sequence of three-way handshake and cleanup. We mark the states ``closed'', ``listen'', and ``established'' in the automata of the peers, because the peer should not stay in other states during communication, based on~\cite{rfc793}. We also mark all states in the automata of the channels and network, to prevent these automata from marking the system. Namely,
\[
    X_{C1,m} = X_{C1}, \quad X_{C2,m} = X_{C2}, \quad X_{C3,m} = X_{C3}, \quad X_{C4,m} = X_{C4}, \quad X_N = X_{N,m}.
\]

\subsection{Safety property models}\label{subsec:tcp:safety}

In~\cite{VonHippel2020}, the safety/liveness property of interest is defined as a threat model (TM). TM explains the property using Linear Temporal Logic (LTL)~\citep{baier_principles_2008}. In this paper, we represent the required properties in~\cite{VonHippel2020} as finite-state automata.

\cite{VonHippel2020} provides one threat model, TM1, for one relevant safety property of TCP. TM1 defines the safety property that if Peer A is at state ``closed'', then Peer B should not be at state ``established'', because both peers should consecutively reach their ``established'' states after beginning the connection handshake. Let $G_{sm}^{TM1}$ be the safety monitor to capture the violation of TM1. We represent $G_{sm}^{TM1}$ as the parallel composition of the automata in~\cref{fig:tcp:peers:timeout} where the marked states are only ``closed'' in Peer A and ``established'' in Peer B, namely $G_{sm}^{TM1} = G_{PA}^{TM1} \parallel G_{PB}^{TM1}$, where $G_{PA}^{TM1} = (X_{PA}^{TM1}, E_{PA}^{TM1}, f_{PA}^{TM1}, x_{PA,0}^{TM1}, X_{PA,m}^{TM1})$ and $G_{PB}^{TM1} = (X_{PB}^{TM1}, E_{PB}^{TM1}, f_{PB}^{TM1}, x_{PB,0}^{TM1}, X_{PB,m}^{TM1})$. Note that
\begin{gather*}
    X_{PA}^{TM1} = X_{PA}, \quad E_{PA}^{TM1} = E_{PA}, \quad x_{PA,0}^{TM1} = x_{PA,0}, \quad X_{PA,m}^{TM1} = \{closed\} \neq X_{PA,m}, \\
    X_{PB}^{TM1} = X_{PB}, \quad E_{PB}^{TM1} = E_{PB}, \quad x_{PB,0}^{TM1} = x_{PB,0}, \quad X_{PB,m}^{TM1} = \{established\} \neq X_{PB,m}
\end{gather*}
Hence, the marked states in $G_{sm}^{TM1}$ are illegal states, capturing that Peer A is at ``closed'' and Peer B is at ``established'' simultaneously.

Since the safety monitor for TM1, $G_{sm}^{TM1}$, is derived from $G_{PA}$ and $G_{PB}$, $G_{other}$ in \cref{alg:attack-spec:safety} is the parallel composition of the automata of the channels and network, namely $G_{other} = G_{C1} \parallel G_{C2} \parallel G_{C3} \parallel G_{C4} \parallel G_N$. Let $H_{nom}$ in \cref{alg:attack-spec:safety} be a nominal specification automaton (without attacker) for TM1. In our system model of TCP, $H_{nom} = Trim(G_{nm}^{TM1} \parallel G_{other})$ has no marked states, thus we conclude that our TCP model, without attackers, is correct in terms of TM1.

\subsection{Nonblockingness property models}\label{subsec:tcp:liveness}

\cite{VonHippel2020} also provides two liveness properties denoted as TM2 and TM3. TM2 defines the liveness property that Peer 2 should eventually reach the ``established'' state. TM3 requires that both peers should not get stuck except at ``closed'' state, that is, no deadlocks except at ``closed'' state are allowed. Both TM2 and TM3 requires the system to remain alive during the communication process. In our case study, we translate TM2 and TM3 into ``equivalent'' nonblockingness properties as expressible representations in the SCT framework, thus slightly abusing the notations ``TM2'' and ``TM3'' in \cite{VonHippel2020}.

We construct the nonblockingness monitors of TM2 and TM3, $G_{nm}^{TM2}$ and $G_{nm}^{TM3}$, by following~\cref{subsec:liveness}. In this case, the nonblockingness monitors are not given as dedicated automata, thus we construct $G_{nm}^{TM2}$ and $G_{nm}^{TM3}$ based on $G_{nom}$ and $G_a$. We discuss the construction of $G_{nm}^{TM2}$ and $G_{nm}^{TM3}$ in~\cref{subsec:tcp:exam}, because to build these automata, we rebuild $G_{nom}$ and $G_a$ as new automata according to TM2 and TM3.

\subsection{Attack model}\label{subsec:tcp:attack-model}

In this section, we explain the attack model for TCP. As we consider the system architecture in~\cref{fig:overview:net} for TCP, the attacker infiltrates the network. First, we construct a modified model of the plant $G_a$ in~\cref{eq:plant:attack} under attack. Since the network of TCP is under attack, we enhance $G_N$ to that under attack, $G_{N,a} = (X_{N,a}, E_{N,a}, f_{N,a}, x_{N,a,0}, X_{N,a,m})$, by adding new transitions and events to represent the capabilities of the attacker. Thus,
\begin{equation}\label{eq:tcp:system:attacked}
    G_a = G_{PA} \parallel G_{PB} \parallel G_{C1} \parallel G_{C2} \parallel G_{C3} \parallel G_{C4} \parallel G_{N,a}
\end{equation}

\cref{fig:tcp:network:attack} depicts the PITM attacked model of the network,
$G_{N,a}$, where ``$ATTK$'' is the set of events of outgoing packets from the
network, namely
\begin{equation}\label{eq:tcp:eventset:attk}
    ATTK = \{SYN_{NC2}, ACK_{NC2}, FIN_{NC2}, SYN\_ACK_{NC2}, SYN_{NC4}, ACK_{NC4}, FIN_{NC4}, SYN\_ACK_{NC4}\},
\end{equation}
representing multiple transitions, illustrated as the red transitions, by
events in $ATTK$. Hence, the event set of $G_{N,a}$, $E_{N,a}$, is as
follows:
\begin{equation}\label{eq:tcp:eventset:network}
    E_{N,a} = ATTK \cup E_N
\end{equation}
where $E_N$ is in~\cref{eq:tcp:events:network:nominal}. This allows the attacker to be flexible so that the attacker can send any packets and freely choose the destination of packets. As in the discussion in~\cref{subsec:models:system} and in the ABP model, we suppose that the attacker cannot control and observe events outside the network. Hence, the event set of $G_a$, $E_a$, is partitioned for controllability and observability of the attacker as follows:
\begin{itemize}
    \item Controllable events: $E_{a,c} = ATTK$ in~\cref{eq:tcp:eventset:attk}
    \item Uncontrollable events: $E_{a,uc} = E_{nom} \setminus E_{a,c}$
    \item Observable events: $E_{a,o} = E_{N,a}$ in~\cref{eq:tcp:eventset:network}
    \item Unobservable events: $E_{a,uo} = E_{nom} \setminus E_{a,o}$
\end{itemize}
In our attack model, the attacker controls the outgoing packets from the network, to lead the safety/nonblockingness monitor to reach its marked (illegal) state.

\subsection{Examination of the PITM attack for TCP}\label{subsec:tcp:exam}

In this section, we examine whether a \textsc{For-all} attack exists in terms of TM1, TM2, and TM3. As in \cref{subsec:abp:exam}, we try to synthesize a \textsc{For-all} attack by the following procedure:
\begin{enumerate}
    \item Construct the plant under attack $G_a$ in \cref{eq:tcp:system:attacked}.
    \item Using \cref{alg:synthesis}, compute the realization of a \textsc{For-all} attack-supervisor with respect to $G_{nom}$, $G_a$ and the safety/nonblockingness monitor for TCP.
\end{enumerate}
As done in \cref{subsec:abp:exam}, if \cref{alg:synthesis} returns the realization, we pick one example string from the initial state to one marked state in $\supcn{\mlang(H_a)}$, which represents one system behaviour under attack that reaches the marked state in the monitor.

\subsubsection{Threat Model 1 with channels}\label{subsubsec:tcp:tm1}

\paragraph{Setup 1}

Let us consider a powerful attacker represented by $G_{N,a}$ in~\cref{fig:tcp:network:attack}. By following the above procedure, $G_a$ has 118761 states and 6307 marked states, and $H_a$ has 38270 states and 704 marked states.

Next, we compute $H_a^{CN}$ with respect to $G_a$ and $H_a$ by following the procedure for TM1. As a result, $H_a^{CN}$ is non-empty, having 52783 states and 626 marked states, and $\mlang(H_a^{CN})$ contains the string
\[ SYN_{BC3}.SYN_{C3N}.SYN\_ACK_{NC4}.SYN\_ACK_{C4B}.ACK_{BC3} \]
which steers $G_{sm}^{TM1}$ to its marked states. Therefore, we conclude that there exists a \textsc{For-all} attacker $S_P$ defined in \cref{eq:marking-supervisor} with respect to $G_a$ and $H_a$ in this setup. From $\lang(G_a) \neq \lang(H_a^{CN})$, the attacker disables some transitions by controllable events in $G_a$, to always eventually win.

\subsubsection{Threat Model 1 without channels}

\paragraph{Setup 2}

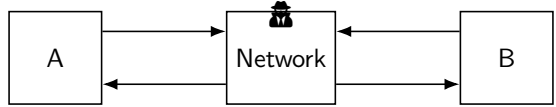
\begin{figure}[htp]
    \centering
    \begin{tikzpicture}[semithick, ->, >=Latex]
    \sffamily
    \node[peer] (peer_a) at (0, 1em) {A};
    \node[peer] (net) at (3cm, 1em) {Network};
    \node[peer] (peer_b) at (6cm, 1em) {B};
    \node[above=8pt of net.center] (evil) {\smilingimp};

    \draw ($(peer_a.east)+(0,1em)$) -- ($(net.west)+(0,1em)$);
    \draw ($(net.west)-(0,1em)$) -- ($(peer_a.east)-(0,1em)$);
    \draw ($(peer_b.west)+(0,1em)$) -- ($(net.east)+(0,1em)$);
    \draw ($(net.east)-(0,1em)$) -- ($(peer_b.west)-(0,1em)$);
\end{tikzpicture}
     \caption{Communication overview without channels}
    \label{fig:overview:nochannels}
\end{figure}

One may find that in our TCP model, the channels just relay the incoming packets to their destinations, without any deletion or manipulation of packets. Since we assume ideal channels, we can reduce the communication architecture in~\cref{fig:overview:net} to that without channels, namely the architecture in~\cref{fig:overview:nochannels}.
Due to the removal of the channels, to assure the synchronization of the peers and network in the parallel composition, we rename the subscripts of the events in $E_{PA}$ in~\cref{eq:tcp:events:peerA}, $E_{PB}$ in~\cref{eq:tcp:events:peerB}, $E_N$ in~\cref{eq:tcp:events:network:nominal}, and $E_{N,a}$ in~\cref{eq:tcp:eventset:network}, as follows:
\begin{equation}\label{eq:tcp:labelchange}
    \begin{gathered}
        AC1 \to AN, \quad C2A \to NA, \quad BC3 \to BN, \quad C4B \to NB, \\
        C1N \to AN, \quad NC2 \to NA, \quad C3N \to BN, \quad NC4 \to NB
    \end{gathered}
\end{equation}
According to this change, the new $G_{nom}$ and $G_a$ are as follows:
\begin{align}
    G_{nom} &=  G_{PA} \parallel G_{PB} \parallel G_N \label{eq:tcp:system:nominal:nochannels} \\
    G_a &= G_{PA} \parallel G_{PB} \parallel G_{N,a} \label{eq:tcp:system:attacked:nochannels}
\end{align}
$G_{nom}$ in~\cref{eq:tcp:system:nominal:nochannels} is trim, consisting of 41 states and 5 marked states, and $G_a$ in~\cref{eq:tcp:system:attacked:nochannels} comprises 580 states and 27 marked states, and is not trim. Since we removed the automata of the channels from our system model, $G_{other}$ and $G_{other,a}$ in~\cref{alg:attack-spec:safety} are equal to $G_N$ and $G_{N,a}$, respectively. Even after the removal of the channels, $H_{nom}$ has no marked states.

Noting that $E_{a,c} \subseteq E_{a,o}$ still holds after renaming, let us revisit the procedure at the beginning of \cref{subsec:tcp:exam} for the construction of $H_a$ and the computation of $H_a^{CN}$ with the new $G_a$. In this setup, $H_a$ consists of 547 states and 3 marked states, $H_a^{CN}$ with respect to $G_a$ and $H_a$ is non-empty with 513 states and 3 marked states. $\mlang(H_a^{CN})$ contains the following string:
\begin{equation}
    SYN_{BN}.SYN_{NB}.ACK_{BN}.ACK_{NB}
\end{equation}
where $SYN_{NB}$ and $ACK_{NB}$ are fake packets inserted by the attacker, tricking Peer B into reaching ``established'' whereas Peer A does not move out from ``closed''. Finally, from $\lang(H_a^{CN}) \neq \lang(G_a)$ and non-trim $G_a$, the attack-supervisor disables several transitions in $G_a$.

\paragraph{Setup 3}

As we have done in the ABP case study, let us consider a less-powerful attacker than the previous setups. First, we change the controllable events for the attacker, $E_{a,c}$, as follows:
\begin{align}
    E_{a,c} &= \{SYN_{AN}, SYN\_ACK_{NB}\} \label{eq:tcp:events:controllable:setup5} \\
    E_{a,uc} &= E_a \setminus E_{a,c}
\end{align}
whereas $E_{a,o}$ and $E_{a,uo}$ do not change. Note that $E_{a,c} \subseteq E_{a,o}$ still holds. $SYN_{AN}$ in $E_{a,c}$ means that the attacker can discard SYN packets coming from Peer A. Next, we redesign the infiltrated network by the attacker, $G_{N,a}$, to represent the reduced capability of the attacker. \cref{fig:tcp:network:attack:less} indicates the model of an infiltrated network by a less powerful attacker, $G_{N,a}^w$. The red transitions are where the attacker can take action.

From the change of $G_{N,a}$ to $G_{N,a}^w$, we change $G_a$ to the entire system under the less powerful PITM attack, namely $G_a = G_{PA} \parallel G_{PB} \parallel G_{N,a}^w$, in this setup. As a result, the new $G_a$ is not trim, consisting of 48 states, 7 marked states, and 1 deadlock state. Because $G_{sm}^{TM1}$ is not different from Setup 2, $G_{other,a} = G_{N,a}^w$ here. Therefore by following the same procedure as above, $H_a$ comprises 47 states and 1 marked state, and $H_a^{CN}$ with respect to $G_a$ and $H_a$ here is non-empty with 63 states and 2 marked states, containing the following string leading $G_{sm}^{TM1}$ to its marked state:
\begin{equation}
    SYN_{BN}.SYN\_ACK_{NB}.ACK_{BN}
\end{equation}
In conclusion, there still exists a \textsc{For-all} attacker with the less-powerful PITM model.

From $G_{N,a}^w$ in~\cref{fig:tcp:network:attack:less}, the attacker can send a fake SYN\_ACK packet to Peer B only when Peer B enters ``SYN sent'' state, and the attacker must keep Peer A at ``closed'' state. Hence, the attacker must disable $SYN_{AN}$ at ``closed'' state in $G_{PA}$ shown in~\cref{fig:tcp:peers:timeout} where the subscripts of events are changed as in~\cref{eq:tcp:labelchange}, and $\lang(H_a^{CN}) \neq \lang(G_a)$ reflects this disablement action. Therefore, if $SYN_{AN}$ is uncontrollable, then $H_a^{CN}$ is empty.

\subsubsection{Threat Model 2}\label{subsubsec:tcp:tm2}

Consider $G_{PA}$, $G_{PB}$, $G_N$, and $G_{N,a}$ in Setup 2. Recall that Threat Model 2 (TM2) requires Peer A to reach its ``established'' state eventually. To design the nonblockingness monitor which captures the violation of TM2, we first unmark all states of $G_{PA}$ and mark its ``established'' state. Let $G_{PA}^{TM2}$ be a new automaton derived from $G_{PA}$ in~\cref{fig:tcp:peers:timeout:A} by this marking and renaming as in~\cref{eq:tcp:labelchange}. In contrast to the construction of safety monitors, $G_{PA}^{TM2}$ captures the desired behaviour where Peer~A reaches its ``established'' state eventually. Thus we construct $G_{nom}$ and $G_a$ as follows:
\begin{align}
    G_{nom} &= G_{PA}^{TM2} \parallel G_{PB} \parallel G_N \label{eq:tcp:system:nominal:tm2} \\
    G_a &= G_{PA}^{TM2} \parallel G_{PB} \parallel G_{N,a} \label{eq:tcp:system:attacked:tm2}
\end{align}
To prevent it from marking $G_{nom}$ and $G_a$, we mark all states in
$G_{PB}$, so the marked states of $G_{nom}$ and $G_a$ are determined by the
``established'' state in $G_{PA}^{TM2}$.

\paragraph{Setup 4}

Let us construct $H_a$ by following \cref{alg:attack-spec:liveness}. First of all, $G_{nom}$ in~\cref{eq:tcp:system:nominal:tm2} is trim, thus the system model without attacker is correct in terms of TM2, meaning that Peer~A eventually reaches its ``established'' state. So, let us proceed to the next step. From the additional transitions of $G_{N,a}$ in~\cref{fig:tcp:network:attack}, $G_a$ in~\cref{eq:tcp:system:attacked:tm2} is not trim, thus $G_a$ contains several deadlock and/or livelock states. In this scenario, we build $G_{nm}^{TM2}$ for TM2 based on $G_a$ and not as a separate automaton. In $G_a$, there are 25 deadlock states. These deadlock states are those the attacker wants $G_a$ to reach so that Peer~A cannot always reach its ``established'' state. To design $G_{nm}^{TM2}$ representing the violation of TM2, namely reaching the deadlock states, we unmark all states in $G_a$ and then mark all the deadlock states. Hence, let $G_{nm}^{TM2}$ be the new automaton built by the marking of deadlock states in $G_a$, so that every string in $\mlang(G_{nm}^{TM2})$ ends with one of the deadlock states in $G_a$. Finally, the specification automaton for the attacker is $H_a = Trim(G_{nm}^{TM2})$.

In this case, $H_a$ consists of 580 states and 25 deadlock states which are determined by $G_a$, and $H_a^{CN}$ with respect to $G_a$ and $H_a$ is non-empty, where $\mlang(H_a^{CN})$ contains the following string:
\begin{equation}\label{eq:tcp:string:setup6}
    SYN_{AN}.SYN\_ACK_{NA}.ACK_{AN}.FIN_{NA}.SYN_{BN}.SYN_{NB}.ACK_{AN}
\end{equation}
$SYN\_ACK_{NA}$, $FIN_{NA}$, and $SYN_{NB}$ in~\cref{eq:tcp:string:setup6} are fake packets inserted by the attacker. This string makes Peer A and Peer B stuck at ``close wait'' state and at ``i1'' state, respectively. Here, $\lang(H_a^{CN}) = \lang(G_a)$, thus the attacker just inserts fake packets and does not disable any controllable events. In conclusion, there exists a \textsc{For-all} attack for TM2 in this setup.

\subsubsection{Threat Model 3}\label{subsusbec:tcp:tm3}

In this section, we examine whether any \textsc{For-all} attacks against the Threat Model 3 (TM3) exist. TM3 captures the following nonblockingness requirement for the system: the peers should not suffer from any deadlocks if they leave ``closed'' state.

Consider $G_{PA}$, $G_{PB}$, $G_N$, and $G_{N,a}$ in Setup 2 again. Since TM3 is defined by a nonblockingness property, we design a nonblockingness monitor for TM3 similarly as a monitor for TM2, discussed in~\cref{subsubsec:tcp:tm2}. According to TM3, we first unmark all states and mark ``closed'' state in $G_{PA}$ and $G_{PB}$. Let $G_{PA}^{TM3}$ and $G_{PB}^{TM3}$ be the new automata derived from $G_{PA}$ and $G_{PB}$ in~\cref{fig:tcp:peers:timeout} by this marking and renaming as in~\cref{eq:tcp:labelchange}, respectively. Since $G_{PA}^{TM3}$ and $G_{PB}^{TM3}$ capture the desired behaviour of the system model, we construct $G_{nom}$ and $G_a$ as follows:
\begin{align}
    G_{nom} &= G_{PA}^{TM3} \parallel G_{PB}^{TM3} \parallel G_N \label{eq:tcp:system:nominal:tm3} \\
    G_a &= G_{PA}^{TM3} \parallel  G_{PB}^{TM3} \parallel G_{N,a} \label{eq:tcp:system:attacked:tm3}
\end{align}
Since all states in $G_N$ and $G_{N,a}$ are marked, the marked states in $G_{nom}$ and $G_a$ are determined by ``closed'' state of $G_{PA}^{TM3}$ and $G_{PB}^{TM3}$.

\paragraph{Setup 5}

We construct $H_a$ using \cref{alg:attack-spec:liveness}. First, $G_{nom}$ in~\cref{eq:tcp:system:nominal:tm3} consisting of 41 states and 1 marked state is trim, thus our system model without attacker is correct in terms of TM3. This means that neither Peer A nor Peer B suffers from deadlocks and/or livelocks when they are not at ``closed'' state. In the next step, due to $G_{N,a}$, $G_a$ in~\cref{eq:tcp:system:attacked:tm3} comprising 580 states and 3 marked states is not trim, thus $G_a$ contains deadlock and/or livelock states. In particular, $G_a$ has 25 deadlock states and no livelock states. Since the nonblockingness monitor for TM3, $G_{nm}^{TM3}$, is not given as a dedicated automaton, $G_{nm}^{TM3}$ is derived from $G_a$ by unmarking all states and marking the 25 deadlock states in $G_a$. Finally, we have $H_a = Trim(G_{nm}^{TM3})$.

As a result, $H_a$ in this setup consists of 580 states and 25 marked (deadlock in $G_a$) states, and $H_a^{CN}$ with respect to $G_a$ and $H_a$ is non-empty with 660 states and 25 marked states. To see a behaviour of the system under the attack, we pick the following example string in $\mlang(H_a^{CN})$:
\begin{equation}\label{eq:tcp:string:setup7}
    listen_A.SYN_{BN}.SYN_{NA}.SYN\_ACK_{AN}.ACK_{NA}.FIN_{AN}.ACK_{NA}
\end{equation}
where the fifth and seventh $ACK_{NA}$ are fake packets sent from the attacker to Peer A. This string makes Peer A and Peer B stuck at ``FIN wait 2'' and ``SYN sent'', respectively. Here, $\lang(H_a^{CN}) = \lang(G_a)$, thus the attacker inserts fake packets and does not disable any controllable events. To sum up, there exists a \textsc{For-all} attack for TM3 in this setup.

\section{Conclusion}\label{sec:conclusions}

We investigated the synthesis problem of \textsc{For-all} attacks under which the attacker can always eventually win, in the specific context of person-in-the-middle attacks on two well-known communication protocols, ABP and TCP, where in each case a sender and a receiver communicate over channels and a network. We formulated this problem in the framework of discrete event systems in order to leverage its supervisory control theory for attacker synthesis. We showed that the synthesis of a \textsc{For-all} attack can be formulated as the problem of finding a maximal controllable and observable sublanguage of the specification language for the attacker with respect to the given plant and the capabilities of the attacker in terms of controllable and observable events. The plant is the combination of the models of the sender, receiver, channels, and network. The specification language for the attacker is derived from a suitable specification automaton; we described in~\cref{subsec:safety,subsec:liveness} how to construct that automaton for various examples of safety properties and nonblockingness properties, respectively. The goal of the attacker is to force a violation of the given safety or nonblockingness property of the communication protocol. We formally derived in~\cref{sec:abp,sec:tcp}, when they existed, several \textsc{For-all} person-in-the-middle attacks for ABP and TCP  under different scenarios of attacker capabilities and safety or nonblockingness property to be violated. We are not aware of any prior work where formal methods are used to synthesize attacks on ABP. For the case of TCP, our results extend  the results in \cite{VonHippel2020}, where the authors considered the synthesis of \textsc{There-exists} attacks under which the attacker may not always win, but will sometimes win.
In total, we presented four setups for ABP and five setups for TCP, where the plant, specification, and event partitions vary. Further setups are discussed in the expanded version of this paper available at \cite{matsui2022synthesis}.

In the PITM attack setups we considered, it was reasonable to assume that the attacker observes all the events it controls. Hence, the synthesis of a \textsc{For-all} attack reduced to the computation of the supremal controllable and normal sublanguage in supervisory control theory of discrete event systems.
This means that the methodology that we employed for ABP and TCP could be applied to other protocols and other types of attacks that can be modelled as additional transitions in the transition structure of the protocol. This shows that formulating attacker synthesis as a supervisory control problem is a powerful approach in the study of vulnerabilities of distributed protocols. In the future, it would be of interest to investigate how to make distributed protocols more resilient to both \textsc{There-exists} and \textsc{For-all} attacks.

\section*{Acknowledgement}

This research was supported in part by the US NSF under grant CNS-1801342. We thank the reviewers for their pertinent comments that helped to improve the presentation of our results.

    \clearpage
    \appendix
\section{Figures of ABP}\label{sec:fig:abp}

\begin{figure}[H]
    \centering
    \subcaptionbox{ABP sender $G_S$\label{fig:abp:models:sender}}[0.6\linewidth]{
        \adjustbox{width=\linewidth}{
            \begin{tikzpicture}[automata]
\node[state, accepting, initial] (s0) {$s_0$};
\node[state, accepting, above right=6mm and 17mm of s0] (s1) {$s_1$};
\node[state, accepting, right=of s1] (s2) {$s_2$};
\node[state, accepting, right=of s2] (s3) {$s_3$};
\node[state, accepting, below right=6mm and 17mm of s3] (s4) {$s_4$};
\node[state, accepting, below left=6mm and 17mm of s4] (s5) {$s_5$};
\node[state, accepting, left=of s5] (s6) {$s_6$};
\node[state, accepting, left=of s6] (s7) {$s_7$};

\path[bend angle=30]
(s0) edge[loop above] node {$a_1'$} ()
(s0) edge[loop below] node {$timeout$} ()
(s0) edge node[xshift=8,yshift=3] {$send$} (s1)
(s1) edge[bend left] node {$p_0$} (s2)
(s2) edge[bend left] node {$timeout$} (s1)
(s2) edge[loop above] node {$a_1'$} ()
(s2) edge node {$a_0'$} (s3)
(s3) edge node[xshift=-8,yshift=3] {$done$} (s4)
(s4) edge[loop above] node {$a_0'$} ()
(s4) edge[loop below] node {$timeout$} ()
(s4) edge[swap] node {$send$} (s5)
(s5) edge[swap, bend right] node {$p_1$} (s6)
(s6) edge[swap, bend right] node {$timeout$} (s5)
(s6) edge[loop below] node {$a_0'$} ()
(s6) edge[swap] node {$a_1'$} (s7)
(s7) edge[swap] node {$done$} (s0)
;
\end{tikzpicture}
         }
    }
    \subcaptionbox{ABP receiver $G_R$\label{Fig:abp:models:receiver}}[0.39\linewidth]{
        \adjustbox{width=\linewidth}{
            \begin{tikzpicture}[automata]
\node[state, accepting, initial] (r0) {$r_0$};
\node[state, accepting, right=of r0] (r1) {$r_1$};
\node[state, accepting, right=of r1] (r2) {$r_2$};
\node[state, accepting, below=15mm of r2] (r3) {$r_3$};
\node[state, accepting, left=of r3] (r4) {$r_4$};
\node[state, accepting, left=of r4] (r5) {$r_5$};

\path[bend angle=20]
(r0) edge node {$p_0'$} (r1)
(r1) edge node {$deliver$} (r2)
(r2) edge[bend left] node {$a_0$} (r3)
(r3) edge[bend left] node {$p_0'$} (r2)
(r3) edge node {$p_1'$} (r4)
(r4) edge[swap] node {$deliver$} (r0)
(r0) edge[bend left] node {$p_1'$} (r5)
(r5) edge[bend left] node {$a_1$} (r0)
;
\end{tikzpicture}
         }
    } \\ \medskip
    \begin{minipage}[c]{0.33\linewidth}
        \centering
        \adjustbox{}{
            \begin{tikzpicture}[automata]
\node[state, accepting, initial] (f0) {$f_0$};
\node[state, accepting, above right=of f0] (f1) {$f_1$};
\node[state, accepting, below right=of f0] (f2) {$f_2$};

\path[bend angle=10]
(f0) edge[loop above] node {$p_0$} ()
(f0) edge[loop below] node {$p_1$} ()
(f0) edge[bend left] node {$p_0$} (f1)
(f1) edge[loop above] node {$p_0$} ()
(f1) edge[loop right] node {$p_0'$} ()
(f1) edge[loop below] node {$p_1$} ()
(f1) edge[bend left] node {$p_0'$} (f0)
(f0) edge[bend left] node {$p_1$} (f2)
(f2) edge[loop above] node {$p_0$} ()
(f2) edge[loop right] node {$p_1'$} ()
(f2) edge[loop below] node {$p_1$} ()
(f2) edge[bend left] node {$p_1'$} (f0)
;
\end{tikzpicture}
         }
        \subcaption{Forward channel $G_{FC}^{nd}$\label{fig:abp:models:forward}}
    \end{minipage}
    \begin{minipage}[c]{0.33\linewidth}
        \centering
        \adjustbox{}{
            \begin{tikzpicture}[automata]
\node[state, accepting, initial] (b0) {$b_0$};
\node[state, accepting, above right=of b0] (b1) {$b_1$};
\node[state, accepting, below right=of b0] (b2) {$b_2$};

\path[bend angle=10]
(b0) edge[loop above] node {$a_0$} ()
(b0) edge[loop below] node {$a_1$} ()
(b0) edge[bend left] node {$a_0$} (b1)
(b1) edge[loop above] node {$a_0$} ()
(b1) edge[loop right] node {$a_0'$} ()
(b1) edge[loop below] node {$a_1$} ()
(b1) edge[bend left] node {$a_0'$} (b0)
(b0) edge[bend left] node {$a_1$} (b2)
(b2) edge[loop above] node {$a_0$} ()
(b2) edge[loop right] node {$a_1'$} ()
(b2) edge[loop below] node {$a_1$} ()
(b2) edge[bend left] node {$a_1'$} (b0)
;
\end{tikzpicture}
         }
        \subcaption{Backward channel $G_{BC}^{nd}$\label{fig:abp:models:backward}}
    \end{minipage}%
    \begin{minipage}[c]{0.33\linewidth}
        \centering
        \subcaptionbox{Sending client $G_{SC}$\label{fig:abp:models:sending}}[\columnwidth]{
            \begin{tikzpicture}[automata, bend angle=20]
\node[state, accepting, initial] (sc0) {$s_0^c$};
\node[state, accepting, right of=sc0] (sc1) {$s_1^c$};

\path
(sc0) edge[bend left] node {$send$} (sc1)
(sc1) edge[bend left] node {$done$} (sc0)
;
\end{tikzpicture}
         } \\ \medskip
        \subcaptionbox{Receiving client $G_{RC}$\label{fig:abp:models:receiving}}[\columnwidth]{
            \begin{tikzpicture}[automata, bend angle=20]
\node[state, accepting, initial] (rc0) {$r_0^c$};

\path (rc0) edge[loop above] node {$deliver$} ();
\end{tikzpicture}
         } \\ \medskip
        \subcaptionbox{Timer $G_T$\label{fig:abp:models:timer}}[\columnwidth]{
            \begin{tikzpicture}[automata, bend angle=20]
\node[state, accepting, initial] (t0) {$t_0$};

\path (t0) edge[loop above] node {$timeout$} ();
\end{tikzpicture}
         }
    \end{minipage}
    \caption{Models of ABP components adopted from~\cite{Alur2017}}
    \label{fig:abp:models}
\end{figure}
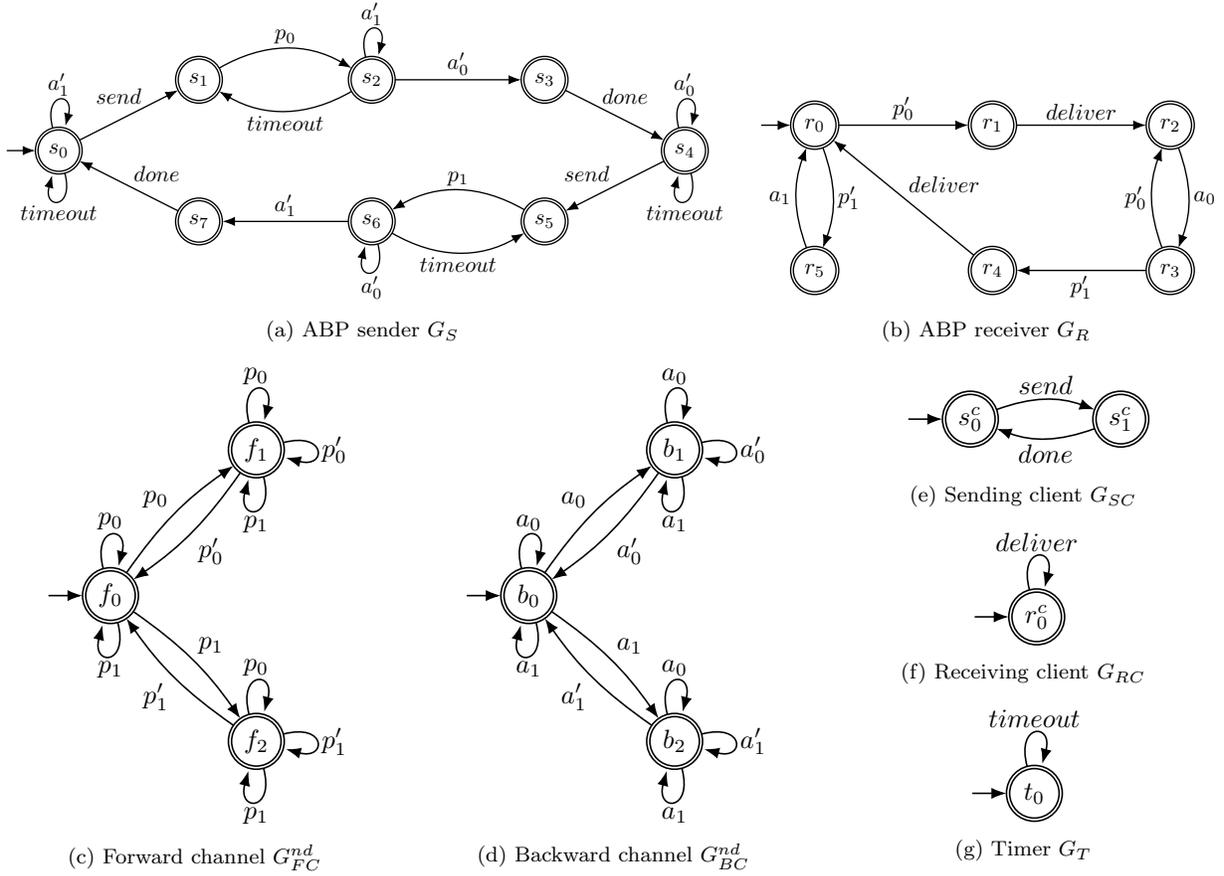

\begin{figure}[htp]
    \centering
    \subcaptionbox{Forward channel $G_{FC}$}[0.49\linewidth]{
        \begin{tikzpicture}[automata, obs]
\node[state, accepting, initial] (f0) {$\{f_0\}$};
\node[state, accepting, above right=of f0] (f0f1) {$\{f_0, f_1\}$};
\node[state, accepting, below right=of f0] (f0f2) {$\{f_0, f_2\}$};
\node[state, accepting, below right=of f0f1] (f0f1f2) {$\{f_0, f_1, f_2\}$};

\path[bend angle=10]
(f0) edge node {$p_0$} (f0f1)
(f0f1) edge[loop above] node {$p_0$} ()
(f0f1) edge[loop below] node {$p_0'$} ()
(f0f1) edge[bend left] node {$p_1$} (f0f1f2)
(f0f1f2) edge[bend left] node {$p_0'$} (f0f1)
(f0) edge node {$p_1$} (f0f2)
(f0f2) edge[loop above] node {$p_1$} ()
(f0f2) edge[loop below] node {$p_1'$} ()
(f0f2) edge[bend left] node {$p_0$} (f0f1f2)
(f0f1f2) edge[bend left] node {$p_1'$} (f0f2)
(f0f1f2) edge[loop above] node {$p_0$} ()
(f0f1f2) edge[loop below] node {$p_1$} ()
;
\end{tikzpicture}
     }
    \subcaptionbox{Backward channel $G_{BC}$}[0.49\linewidth]{
        \begin{tikzpicture}[automata, obs]
\node[state, accepting, initial] (b0) {$\{b_0\}$};
\node[state, accepting, above right=of b0] (b0b1) {$\{b_0, b_1\}$};
\node[state, accepting, below right=of b0] (b0b2) {$\{b_0, b_2\}$};
\node[state, accepting, below right=of b0b1] (b0b1b2) {$\{b_0, b_1, b_2\}$};

\path[bend angle=10]
(b0) edge node {$a_0$} (b0b1)
(b0b1) edge[loop above] node {$a_0$} ()
(b0b1) edge[loop below] node {$a_0'$} ()
(b0b1) edge[bend left] node {$a_1$} (b0b1b2)
(b0b1b2) edge[bend left] node {$a_0'$} (b0b1)
(b0) edge node {$a_1$} (b0b2)
(b0b2) edge[loop above] node {$a_1$} ()
(b0b2) edge[loop below] node {$a_1'$} ()
(b0b2) edge[bend left] node {$a_0$} (b0b1b2)
(b0b1b2) edge[bend left] node {$a_1'$} (b0b2)
(b0b1b2) edge[loop above] node {$a_0$} ()
(b0b1b2) edge[loop below] node {$a_1$} ()
;
\end{tikzpicture}
     }
    \caption{Observer automata of channels}
    \label{fig:abp:obs-channels}
\end{figure}
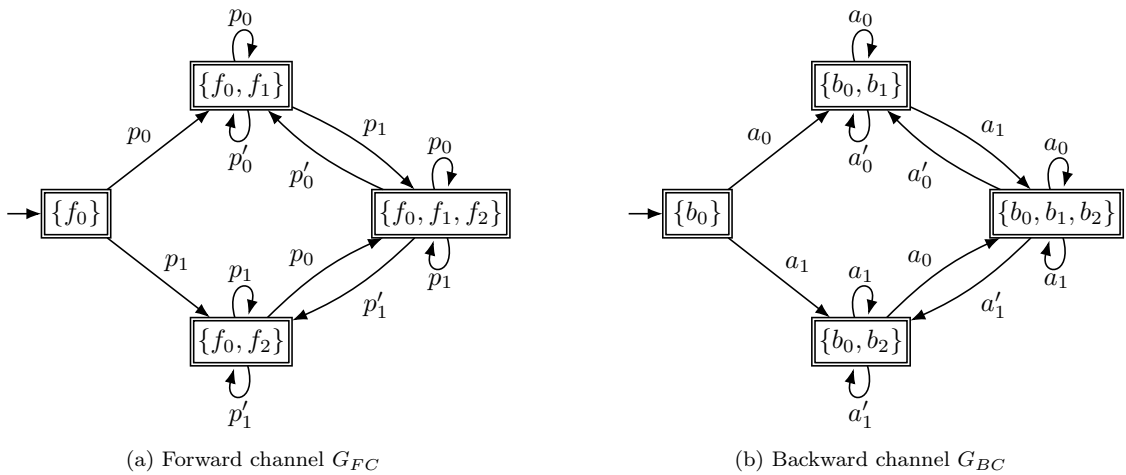

\begin{figure}[htp]
    \centering
    \subcaptionbox{Safety monitor 1 $G_{sm}^1$; $send$ and $deliver$ should happen in the right order.\label{fig:abp:safety:1}}[0.49\linewidth]{
        \begin{tikzpicture}[automata]
\node[state, initial] (q0) {$q_0$};
\node[state, below right=of q0] (q1) {$q_1$};
\node[state, above right=of q1, accepting] (q2) {$q_2$};

\path[bend angle=20]
(q0) edge[bend left] node {$send$} (q1)
(q1) edge[bend left] node {$deliver$} (q0)
(q1) edge[swap, bend right] node {$send$} (q2)
(q0) edge node {$deliver$} (q2)
(q2) edge[loop above] node {$send$} ()
(q2) edge[loop right] node {$deliver$} ()
;
\end{tikzpicture}
     }
    \subcaptionbox{Safety monitor 2 $G_{sm}^2$; $deliver$ and $done$ should happen in the right order.\label{fig:abp:safety:2}}[0.49\linewidth]{
        \begin{tikzpicture}[automata]
\node[state, initial] (q0) {$q_0$};
\node[state, below right=of q0] (q1) {$q_1$};
\node[state, above right=of q1, accepting] (q2) {$q_2$};

\path[bend angle=20]
(q0) edge[bend left] node {$deliver$} (q1)
(q1) edge[bend left] node {$done$} (q0)
(q1) edge[bend right] node {$deliver$} (q2)
(q0) edge node {$done$} (q2)
(q2) edge[loop above] node {$deliver$} ()
(q2) edge[loop right] node {$done$} ()
;
\end{tikzpicture}
     }
    \caption{Safety monitors from~\cite{Alur2017}}
    \label{fig:abp:safety-monitors}
\end{figure}
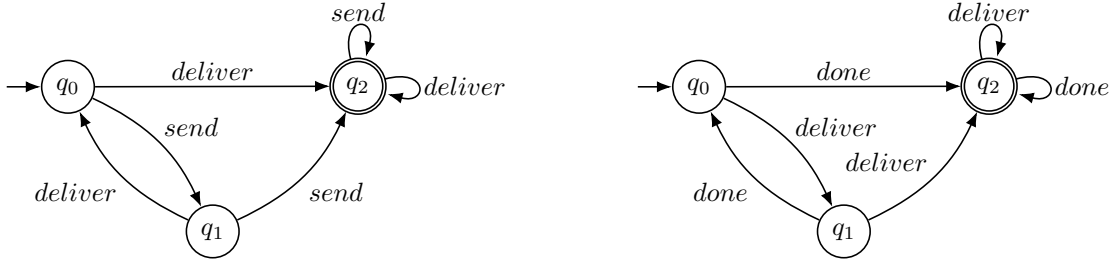

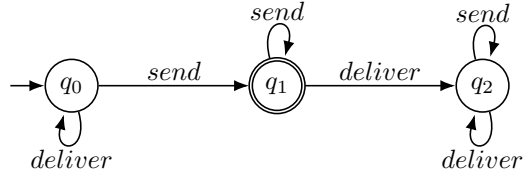
\begin{figure}[htp]
    \centering
    \begin{tikzpicture}[automata]
\node[state, initial] (q0) {$q_0$};
\node[state, right=of q0, accepting] (q1) {$q_1$};
\node[state, right=of q1] (q2) {$q_2$};

\path[bend angle=20]
(q0) edge[loop below] node {$deliver$} ()
(q0) edge node {$send$} (q1)
(q1) edge[loop above] node {$send$} ()
(q1) edge node {$deliver$} (q2)
(q2) edge[loop above] node {$send$} ()
(q2) edge[loop below] node {$deliver$} ()
;
\end{tikzpicture}
     \caption{Nonblockingness monitor $G_{nm}$ inspired by~\cite{Alur2017}; the first $send$ should eventually be followed by a $deliver$}
    \label{fig:abp:liveness-monitor}
\end{figure}

\begin{figure}[htp]
    \centering
    \subcaptionbox{Forward MITM channel $G_{FC,a}^{nd}$\label{fig:abp:mitm:forward}}[0.49\linewidth]{
        \adjustbox{scale=0.9}
{
        \begin{tikzpicture}[automata]
\node[state, accepting, initial] (f0) {$f_0$};
\node[state, accepting, above right=of f0] (f1) {$f_1$};
\node[state, accepting, below right=of f0] (f2) {$f_2$};

\path[bend angle=10]
(f0) edge[loop above] node {$p_0$} ()
(f0) edge[loop below] node {$p_1$} ()
(f0) edge[bend left] node {$p_0$} (f1)
(f1) edge[loop above] node {$p_0$} ()
(f1) edge[loop right] node {$p_0'$} ()
(f1) edge[loop below] node {$p_1$} ()
(f1) edge[bend left] node {$p_0'$} (f0)
(f0) edge[bend left] node {$p_1$} (f2)
(f2) edge[loop above] node {$p_0$} ()
(f2) edge[loop right] node {$p_1'$} ()
(f2) edge[loop below] node {$p_1$} ()
(f2) edge[bend left] node {$p_1'$} (f0)

(f1) edge[red, bend right=80, swap] node {$p_1'$} (f0)
(f1) edge[red, out=-30, in=-60, loop] node {$p_1'$} ()
(f2) edge[red, bend left=80] node {$p_0'$} (f0)
(f2) edge[red, in=30, out=60, loop] node {$p_0'$} ()
;
\end{tikzpicture}
 }
    }
    \subcaptionbox{Backward MITM channel $G_{BC,a}^{nd}$\label{fig:abp:mitm:backward}}[0.49\linewidth]{
        \adjustbox{scale=0.9}
{
        \begin{tikzpicture}[automata]
\node[state, accepting, initial] (b0) {$b_0$};
\node[state, accepting, above right=of b0] (b1) {$b_1$};
\node[state, accepting, below right=of b0] (b2) {$b_2$};

\path
(b0) edge[loop above] node {$a_0$} ()
(b0) edge[loop below] node {$a_1$} ()
(b0) edge[bend left] node {$a_0$} (b1)
(b1) edge[loop above] node {$a_0$} ()
(b1) edge[loop right] node {$a_0'$} ()
(b1) edge[loop below] node {$a_1$} ()
(b1) edge[bend left] node {$a_0'$} (b0)
(b0) edge[bend left] node {$a_1$} (b2)
(b2) edge[loop above] node {$a_0$} ()
(b2) edge[loop right] node {$a_1'$} ()
(b2) edge[loop below] node {$a_1$} ()
(b2) edge[bend left] node {$a_1'$} (b0)

(b1) edge[red, bend right=80, swap] node {$a_1'$} (b0)
(b1) edge[red, out=-30, in=-60, loop] node {$a_1'$} ()
(b2) edge[red, bend left=80] node {$a_0'$} (b0)
(b2) edge[red, in=30, out=60, loop] node {$a_0'$} ()
;
\end{tikzpicture}
 }
    }
    \caption{Channel models under the MITM attack}
    \label{fig:abp:mitm}
\end{figure}
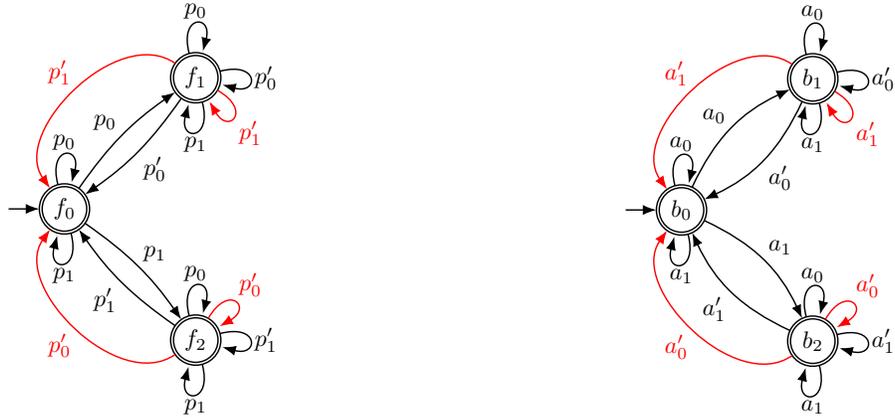

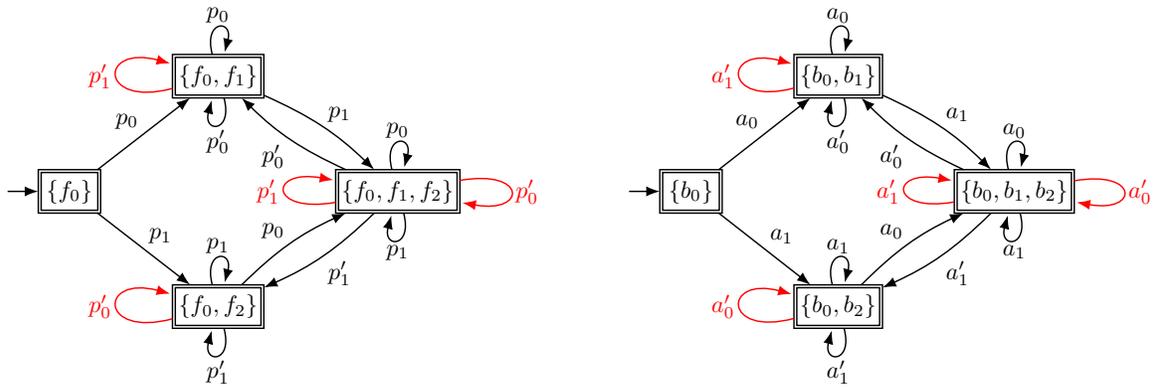
\begin{figure}[htp]
    \centering
    \subcaptionbox{Foward MITM channel $G_{FC,a}$\label{fig:abp:mitm:obs:forward}}[0.49\linewidth]{
        \adjustbox{scale=0.9}
        {
            \begin{tikzpicture}[automata, obs]
\node[state, accepting, initial] (f0) {$\{f_0\}$};
\node[state, accepting, above right=of f0] (f0f1) {$\{f_0, f_1\}$};
\node[state, accepting, below right=of f0] (f0f2) {$\{f_0, f_2\}$};
\node[state, accepting, below right=of f0f1] (f0f1f2) {$\{f_0, f_1, f_2\}$};

\path[bend angle=10]
(f0) edge node {$p_0$} (f0f1)
(f0f1) edge[loop above] node {$p_0$} ()
(f0f1) edge[loop below] node {$p_0'$} ()
(f0f1) edge[bend left] node {$p_1$} (f0f1f2)
(f0f1f2) edge[bend left] node {$p_0'$} (f0f1)
(f0) edge node {$p_1$} (f0f2)
(f0f2) edge[loop above] node {$p_1$} ()
(f0f2) edge[loop below] node {$p_1'$} ()
(f0f2) edge[bend left] node {$p_0$} (f0f1f2)
(f0f1f2) edge[bend left] node {$p_1'$} (f0f2)
(f0f1f2) edge[loop above] node {$p_0$} ()
(f0f1f2) edge[loop below] node {$p_1$} ()

(f0f1) edge[red, loop left] node {$p_1'$} ()
(f0f2) edge[red, loop left] node {$p_0'$} ()
(f0f1f2) edge[red, out=10, in=-10, loop] node {$p_0'$} ()
(f0f1f2) edge[red, out=190, in=170, loop] node {$p_1'$} ()
;
\end{tikzpicture}
         }
    }
    \subcaptionbox{Backward MITM channel $G_{BC,a}$\label{fig:abp:mitm:obs:backward}}[0.49\linewidth]{
        \adjustbox{scale=0.9}
        {
            \begin{tikzpicture}[automata, obs]
\node[state, accepting, initial] (b0) {$\{b_0\}$};
\node[state, accepting, above right=of b0] (b0b1) {$\{b_0, b_1\}$};
\node[state, accepting, below right=of b0] (b0b2) {$\{b_0, b_2\}$};
\node[state, accepting, below right=of b0b1] (b0b1b2) {$\{b_0, b_1, b_2\}$};

\path[bend angle=10]
(b0) edge node {$a_0$} (b0b1)
(b0b1) edge[loop above] node {$a_0$} ()
(b0b1) edge[loop below] node {$a_0'$} ()
(b0b1) edge[bend left] node {$a_1$} (b0b1b2)
(b0b1b2) edge[bend left] node {$a_0'$} (b0b1)
(b0) edge node {$a_1$} (b0b2)
(b0b2) edge[loop above] node {$a_1$} ()
(b0b2) edge[loop below] node {$a_1'$} ()
(b0b2) edge[bend left] node {$a_0$} (b0b1b2)
(b0b1b2) edge[bend left] node {$a_1'$} (b0b2)
(b0b1b2) edge[loop above] node {$a_0$} ()
(b0b1b2) edge[loop below] node {$a_1$} ()

(b0b1) edge[red, loop left] node {$a_1'$} ()
(b0b2) edge[red, loop left] node {$a_0'$} ()
(b0b1b2) edge[red, out=10, in=-10, loop] node {$a_0'$} ()
(b0b1b2) edge[red, out=190, in=170, loop] node {$a_1'$} ()
;
\end{tikzpicture}
         }
    }
    \caption{Observer automata of the MITM channels}
    \label{fig:abp:mitm:obs}
\end{figure}

\begin{figure}[htp]
    \centering
    \begin{minipage}[t]{0.4\linewidth}
        \centering
        \begin{tikzpicture}[automata]
\node[state, accepting, initial] (f0) {$f_0$};
\node[state, accepting, above right=of f0] (f1) {$f_1$};
\node[state, accepting, below right=of f0] (f2) {$f_2$};

\path
(f0) edge[loop above] node {$p_0$} ()
(f0) edge[loop below] node {$p_1$} ()
(f0) edge[bend left] node {$p_0$} (f1)
(f1) edge[loop above] node {$p_0$} ()
(f1) edge[loop right] node {$p_0'$} ()
(f1) edge[loop below] node {$p_1$} ()
(f1) edge[bend left] node {$p_0'$} (f0)
(f0) edge[bend left] node {$p_1$} (f2)
(f2) edge[loop above] node {$p_0$} ()
(f2) edge[loop right] node {$p_1'$} ()
(f2) edge[loop below] node {$p_1$} ()
(f2) edge[bend left] node {$p_1'$} (f0)

(f1) edge[red, bend right=80, swap] node {$p_1'$} (f0)
;
\end{tikzpicture}
         \subcaption{Lesspowerful forward MITM channel $G_{FC,wa}^{nd}$}
        \label{fig:abp:weak:forward}
    \end{minipage}\hfill \begin{minipage}[t]{0.53\linewidth}
        \centering
        \begin{tikzpicture}[automata, obs]
\node[state, accepting, initial] (f0) {$\{f_0\}$};
\node[state, accepting, above right=of f0] (f0f1) {$\{f_0, f_1\}$};
\node[state, accepting, below right=of f0] (f0f2) {$\{f_0, f_2\}$};
\node[state, accepting, below right=of f0f1] (f0f1f2) {$\{f_0, f_1, f_2\}$};

\path[bend angle=10]
(f0) edge[bend left] node {$p_0$} (f0f1)
(f0f1) edge[loop above] node {$p_0$} ()
(f0f1) edge[loop below] node {$p_0'$} ()
(f0f1) edge[bend left, red] node {$p_1'$} (f0)
(f0f1) edge[bend left] node {$p_1$} (f0f1f2)
(f0f1f2) edge[bend left] node {$p_0'$} (f0f1)
(f0) edge node {$p_1$} (f0f2)
(f0f2) edge[loop above] node {$p_1$} ()
(f0f2) edge[loop below] node {$p_1'$} ()
(f0f2) edge[bend left] node {$p_0$} (f0f1f2)
(f0f1f2) edge[bend left] node {$p_1'$} (f0f2)
(f0f1f2) edge[loop above] node {$p_0$} ()
(f0f1f2) edge[loop below] node {$p_1$} ()
;
\end{tikzpicture}
         \adjustbox{phantom=v}{\begin{tikzpicture}[automata]
\node[state, accepting, initial] (f0) {$f_0$};
\node[state, accepting, above right=of f0] (f1) {$f_1$};
\node[state, accepting, below right=of f0] (f2) {$f_2$};

\path
(f0) edge[loop above] node {$p_0$} ()
(f0) edge[loop below] node {$p_1$} ()
(f0) edge[bend left] node {$p_0$} (f1)
(f1) edge[loop above] node {$p_0$} ()
(f1) edge[loop right] node {$p_0'$} ()
(f1) edge[loop below] node {$p_1$} ()
(f1) edge[bend left] node {$p_0'$} (f0)
(f0) edge[bend left] node {$p_1$} (f2)
(f2) edge[loop above] node {$p_0$} ()
(f2) edge[loop right] node {$p_1'$} ()
(f2) edge[loop below] node {$p_1$} ()
(f2) edge[bend left] node {$p_1'$} (f0)

(f1) edge[red, bend right=80, swap] node {$p_1'$} (f0)
;
\end{tikzpicture}
 }
        \subcaption{Observer automata of lesspowerful forward MITM channel $G_{FC,wa}$}
        \label{fig:abp:weak:forward:obs}
    \end{minipage}
    \caption{Lesspowerful forward MITM channel}
    \label{fig:abp:weak}
\end{figure}
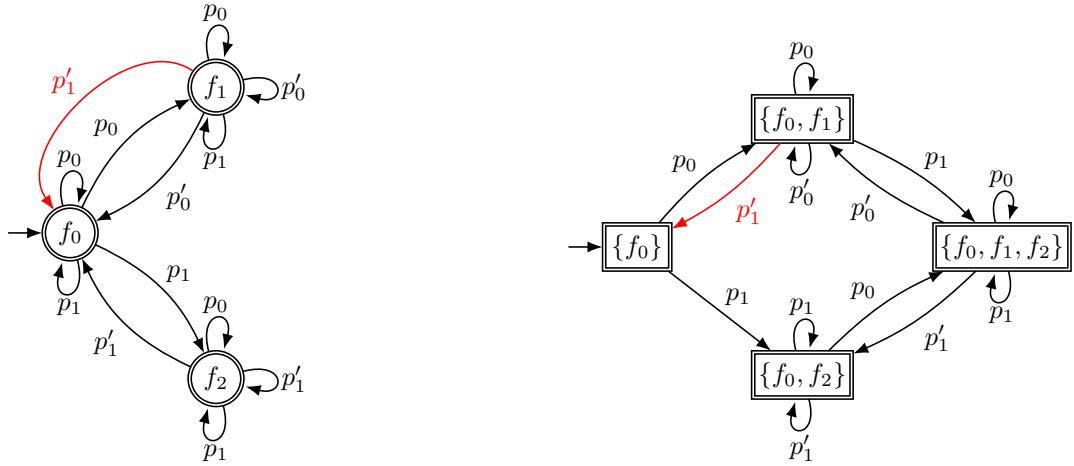

\begin{figure}[htp]
    \centering
    \begin{tikzpicture}[automata]
\node[state, accepting, initial] (f0) {$f_0$};
\node[state, accepting, above right=of f0] (f1) {$f_1$};
\node[state, accepting, below right=of f0] (f2) {$f_2$};
\node[state, accepting, below right=of f1] (f0p) {$f_0'$};
\node[state, accepting, above right=of f0p] (f1p) {$f_1'$};
\node[state, accepting, below right=of f0p] (f2p) {$f_2'$};

\path[bend angle=10]
(f0) edge[loop above] node {$p_0$} ()
(f0) edge[loop below] node {$p_1$} ()
(f0) edge[bend left] node {$p_0$} (f1)
(f1) edge[loop above] node {$p_0$} ()
(f1) edge[loop right] node {$p_0'$} ()
(f1) edge[loop below] node {$p_1$} ()
(f1) edge[bend left] node {$p_0'$} (f0)
(f0) edge[bend left] node {$p_1$} (f2)
(f2) edge[loop above] node {$p_0$} ()
(f2) edge[loop right] node {$p_1'$} ()
(f2) edge[loop below] node {$p_1$} ()
(f2) edge[bend left] node {$p_1'$} (f0)
(f0p) edge[loop above] node {$p_0$} ()
(f0p) edge[loop below] node {$p_1$} ()
(f0p) edge[bend left] node {$p_0$} (f1p)
(f1p) edge[loop above] node {$p_0$} ()
(f1p) edge[loop right] node {$p_0'$} ()
(f1p) edge[loop below] node {$p_1$} ()
(f1p) edge[bend left] node {$p_0'$} (f0p)
(f0p) edge[bend left] node {$p_1$} (f2p)
(f2p) edge[loop above] node {$p_0$} ()
(f2p) edge[loop right] node {$p_1'$} ()
(f2p) edge[loop below] node {$p_1$} ()
(f2p) edge[bend left] node {$p_1'$} (f0p)
(f1) edge[red] node {$p_1'$} (f0p)
;
\end{tikzpicture}
     \caption{One-shot forward MITM channel}\label{fig:abp:mitm:oneshot}
\end{figure}
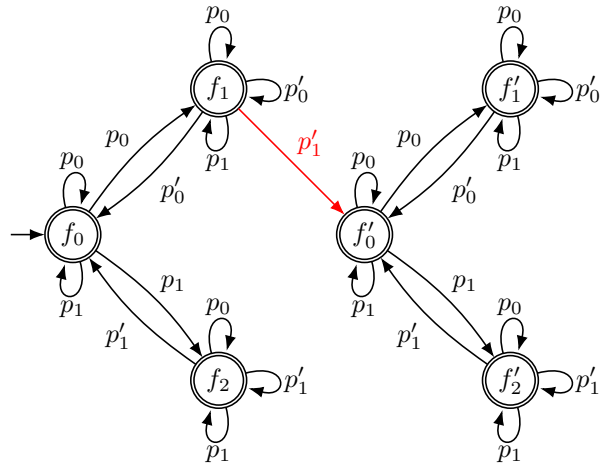

\clearpage

\section{Figures of TCP}\label{sec:fig:tcp}

\begin{figure}[htp]
    \newcommand*{\chscale}{0.6}
    \newcommand*{\chcapscale}{0.245\linewidth}
    \centering
    \subcaptionbox{Channel 1 $G_{C1}$}[\chcapscale]{
        \adjustbox{scale=\chscale}
{
        \begin{tikzpicture}[automata, node distance=20mm]
    \renewcommand*{\pout}{\tss{C1N}}
    \renewcommand*{\pin}{\tss{AC1}}

    \node[state, accepting, initial] (init) {};
    \node[state, accepting, above=of init] (SYN) {};
    \node[state, accepting, right=of init] (ACK) {};
    \node[state, accepting, below=of init] (SYN_ACK) {};
    \node[state, accepting, left=of init] (FIN) {};

    \sffamily
    \path
    (init) edge[bend left] node {SYN\pin} (SYN)
    (SYN) edge[bend left] node {SYN\pout} (init)
    (init) edge[bend left] node {ACK\pin} (ACK)
    (ACK) edge[bend left] node {ACK\pout} (init)
    (init) edge[bend left] node {SYN\_ACK\pin} (SYN_ACK)
    (SYN_ACK) edge[bend left] node {SYN\_ACK\pout} (init)
    (init) edge[bend left] node {FIN\pin} (FIN)
    (FIN) edge[bend left] node {FIN\pout} (init)
    ;
\end{tikzpicture}
 }
    }
    \subcaptionbox{Channel 2 $G_{C2}$}[\chcapscale]{
        \adjustbox{scale=\chscale}
{
        \begin{tikzpicture}[automata, node distance=20mm]
    \renewcommand*{\pout}{\tss{C2A}}
    \renewcommand*{\pin}{\tss{NC2}}

    \node[state, accepting, initial] (init) {};
    \node[state, accepting, above=of init] (SYN) {};
    \node[state, accepting, right=of init] (ACK) {};
    \node[state, accepting, below=of init] (SYN_ACK) {};
    \node[state, accepting, left=of init] (FIN) {};

    \sffamily
    \path
    (init) edge[bend left] node {SYN\pin} (SYN)
    (SYN) edge[bend left] node {SYN\pout} (init)
    (init) edge[bend left] node {ACK\pin} (ACK)
    (ACK) edge[bend left] node {ACK\pout} (init)
    (init) edge[bend left] node {SYN\_ACK\pin} (SYN_ACK)
    (SYN_ACK) edge[bend left] node {SYN\_ACK\pout} (init)
    (init) edge[bend left] node {FIN\pin} (FIN)
    (FIN) edge[bend left] node {FIN\pout} (init)
    ;
\end{tikzpicture}
 }
    }
    \subcaptionbox{Channel 3 $G_{C3}$}[\chcapscale]{
        \adjustbox{scale=\chscale}
{
        \begin{tikzpicture}[automata, node distance=20mm]
    \renewcommand*{\pout}{\tss{C3N}}
    \renewcommand*{\pin}{\tss{BC3}}

    \node[state, accepting, initial] (init) {};
    \node[state, accepting, above=of init] (SYN) {};
    \node[state, accepting, right=of init] (ACK) {};
    \node[state, accepting, below=of init] (SYN_ACK) {};
    \node[state, accepting, left=of init] (FIN) {};

    \sffamily
    \path
    (init) edge[bend left] node {SYN\pin} (SYN)
    (SYN) edge[bend left] node {SYN\pout} (init)
    (init) edge[bend left] node {ACK\pin} (ACK)
    (ACK) edge[bend left] node {ACK\pout} (init)
    (init) edge[bend left] node {SYN\_ACK\pin} (SYN_ACK)
    (SYN_ACK) edge[bend left] node {SYN\_ACK\pout} (init)
    (init) edge[bend left] node {FIN\pin} (FIN)
    (FIN) edge[bend left] node {FIN\pout} (init)
    ;
\end{tikzpicture}
 }
    }
    \subcaptionbox{Channel 4 $G_{C4}$}[\chcapscale]{
        \adjustbox{scale=\chscale}
{
        \begin{tikzpicture}[automata, node distance=20mm]
    \renewcommand*{\pout}{\tss{C4B}}
    \renewcommand*{\pin}{\tss{NC4}}

    \node[state, accepting, initial] (init) {};
    \node[state, accepting, above=of init] (SYN) {};
    \node[state, accepting, right=of init] (ACK) {};
    \node[state, accepting, below=of init] (SYN_ACK) {};
    \node[state, accepting, left=of init] (FIN) {};

    \sffamily
    \path
    (init) edge[bend left] node {SYN\pin} (SYN)
    (SYN) edge[bend left] node {SYN\pout} (init)
    (init) edge[bend left] node {ACK\pin} (ACK)
    (ACK) edge[bend left] node {ACK\pout} (init)
    (init) edge[bend left] node {SYN\_ACK\pin} (SYN_ACK)
    (SYN_ACK) edge[bend left] node {SYN\_ACK\pout} (init)
    (init) edge[bend left] node {FIN\pin} (FIN)
    (FIN) edge[bend left] node {FIN\pout} (init)
    ;
\end{tikzpicture}
 }
    }
    \caption{Channel models of TCP}
    \label{fig:tcp:component:2}
\end{figure}
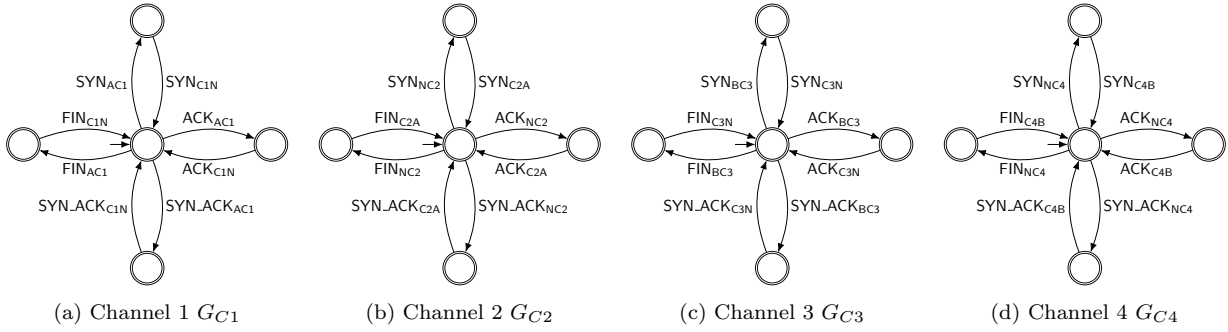

\begin{figure}[htp]
    \centering
    \begin{minipage}[b]{0.49\linewidth}
        \centering
        \adjustbox{height=0.25\textheight}
        {
            \begin{tikzpicture}[automata, >=latex, bend angle=7]
    \newcommand*{\distance}{4}
    \node[state, accepting, initial] (init) at (0,0) {};
    \node[state, accepting] (SYN_btoa) at (22.5:\distance) {};
    \node[state, accepting] (SYN_atob) at (67.5:\distance) {};
    \node[state, accepting] (SYN_ACK_btoa) at (112.5:\distance) {};
    \node[state, accepting] (SYN_ACK_atob) at (157.5:\distance) {};
    \node[state, accepting] (ACK_btoa) at (202.5:\distance) {};
    \node[state, accepting] (ACK_atob) at (247.5:\distance) {};
    \node[state, accepting] (FIN_btoa) at (292.5:\distance) {};
    \node[state, accepting] (FIN_atob) at (337.5:\distance) {};

    \sffamily
    \path
    (init) edge[bend left] node[pos=0.65]{SYN\tss{C1N}} (SYN_atob)
    (SYN_atob) edge[bend left] node[very near start]{SYN\tss{NC4}} (init)
    (init) edge[bend left] node[very near end]{SYN\tss{C3N}} (SYN_btoa)
    (SYN_btoa) edge[bend left] node[near start]{SYN\tss{NC2}} (init)
    (init) edge[bend left] node{FIN\tss{C1N}} (FIN_atob)
    (FIN_atob) edge[bend left] node[very near start]{FIN\tss{NC4}} (init)
    (init) edge[bend left] node[very near end]{FIN\tss{C3N}} (FIN_btoa)
    (FIN_btoa) edge[bend left] node[pos=0.4,xshift=5]{FIN\tss{NC2}} (init)
    (init) edge[bend left] node[pos=0.8]{ACK\tss{C1N}} (ACK_atob)
    (ACK_atob) edge[bend left] node[very near start]{ACK\tss{NC4}} (init)
    (init) edge[bend left] node[very near end]{ACK\tss{C3N}} (ACK_btoa)
    (ACK_btoa) edge[bend left] node[near start]{ACK\tss{NC2}} (init)
    (init) edge[bend left] node[pos=0.65]{SYN\_ACK\tss{C1N}} (SYN_ACK_atob)
    (SYN_ACK_atob) edge[bend left] node[pos=0.05,yshift=-3]{SYN\_ACK\tss{NC4}} (init)
    (init) edge[bend left] node[very near end]{SYN\_ACK\tss{C3N}} (SYN_ACK_btoa)
    (SYN_ACK_btoa) edge[bend left] node[very near start]{SYN\_ACK\tss{NC2}} (init)
    ;
\end{tikzpicture}
         }
        \caption{Network model of TCP\label{fig:tcp:component:3}}
    \end{minipage}
    \begin{minipage}[b]{0.49\linewidth}
        \centering
        \adjustbox{height=0.25\textheight}
        {
            \begin{tikzpicture}[automata, >=latex, bend angle=7]
    \newcommand*{\distance}{4}
    \node[state, accepting, initial] (init) at (0,0) {};
    \node[state, accepting] (SYN_btoa) at (22.5:\distance) {};
    \node[state, accepting] (SYN_atob) at (67.5:\distance) {};
    \node[state, accepting] (SYN_ACK_btoa) at (112.5:\distance) {};
    \node[state, accepting] (SYN_ACK_atob) at (157.5:\distance) {};
    \node[state, accepting] (ACK_btoa) at (202.5:\distance) {};
    \node[state, accepting] (ACK_atob) at (247.5:\distance) {};
    \node[state, accepting] (FIN_btoa) at (292.5:\distance) {};
    \node[state, accepting] (FIN_atob) at (337.5:\distance) {};

    \sffamily
    \path
    (init) edge[bend left] node[pos=0.65]{SYN\tss{C1N}} (SYN_atob)
    (SYN_atob) edge[bend left, red] node[very near start]{ATTK} (init)
    (init) edge[bend left] node[very near end]{SYN\tss{C3N}} (SYN_btoa)
    (SYN_btoa) edge[bend left, red] node[near start]{ATTK} (init)
    (init) edge[bend left] node{FIN\tss{C1N}} (FIN_atob)
    (FIN_atob) edge[bend left, red] node[very near start]{ATTK} (init)
    (init) edge[bend left] node[very near end]{FIN\tss{C3N}} (FIN_btoa)
    (FIN_btoa) edge[bend left, red] node[pos=0.4]{ATTK} (init)
    (init) edge[bend left] node[pos=0.8]{ACK\tss{C1N}} (ACK_atob)
    (ACK_atob) edge[bend left, red] node[very near start]{ATTK} (init)
    (init) edge[bend left] node[very near end]{ACK\tss{C3N}} (ACK_btoa)
    (ACK_btoa) edge[bend left, red] node[near start]{ATTK} (init)
    (init) edge[bend left] node[pos=0.65]{SYN\_ACK\tss{C1N}} (SYN_ACK_atob)
    (SYN_ACK_atob) edge[bend left, red] node[pos=0.05]{ATTK} (init)
    (init) edge[bend left] node[very near end]{SYN\_ACK\tss{C3N}} (SYN_ACK_btoa)
    (SYN_ACK_btoa) edge[bend left, red] node[very near start]{ATTK} (init)
    ;
\end{tikzpicture}
         }
        \caption{Network model under the MITM attack $G_{N,a}$\label{fig:tcp:network:attack}}
    \end{minipage}
\end{figure}

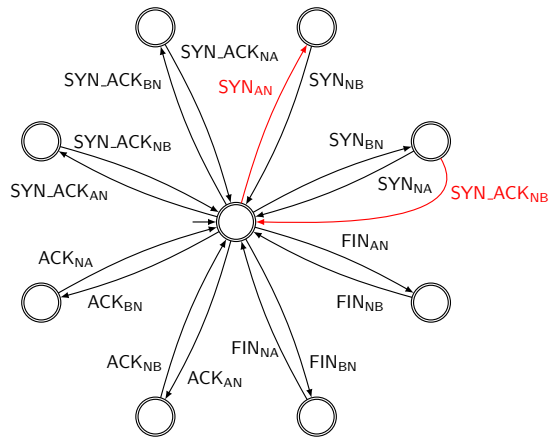
\begin{figure}[H]
    \centering
    \adjustbox{height=0.25\textheight}{
        \begin{tikzpicture}[automata, >=latex, bend angle=7]
    \newcommand*{\distance}{4}
    \node[state, accepting, initial] (init) at (0,0) {};
    \node[state, accepting] (SYN_btoa) at (22.5:\distance) {};
    \node[state, accepting] (SYN_atob) at (67.5:\distance) {};
    \node[state, accepting] (SYN_ACK_btoa) at (112.5:\distance) {};
    \node[state, accepting] (SYN_ACK_atob) at (157.5:\distance) {};
    \node[state, accepting] (ACK_btoa) at (202.5:\distance) {};
    \node[state, accepting] (ACK_atob) at (247.5:\distance) {};
    \node[state, accepting] (FIN_btoa) at (292.5:\distance) {};
    \node[state, accepting] (FIN_atob) at (337.5:\distance) {};

    \sffamily
    \path
    (init) edge[bend left, red] node[pos=0.65]{SYN\tss{AN}} (SYN_atob)
    (SYN_atob) edge[bend left] node[very near start]{SYN\tss{NB}} (init)
    (init) edge[bend left] node[very near end]{SYN\tss{BN}} (SYN_btoa)
    (SYN_btoa) edge[bend left] node[near start]{SYN\tss{NA}} (init)
    (init) edge[bend left] node{FIN\tss{AN}} (FIN_atob)
    (FIN_atob) edge[bend left] node[very near start]{FIN\tss{NB}} (init)
    (init) edge[bend left] node[very near end]{FIN\tss{BN}} (FIN_btoa)
    (FIN_btoa) edge[bend left] node[pos=0.4,xshift=5]{FIN\tss{NA}} (init)
    (init) edge[bend left] node[pos=0.8]{ACK\tss{AN}} (ACK_atob)
    (ACK_atob) edge[bend left] node[very near start]{ACK\tss{NB}} (init)
    (init) edge[bend left] node[very near end]{ACK\tss{BN}} (ACK_btoa)
    (ACK_btoa) edge[bend left] node[near start]{ACK\tss{NA}} (init)
    (init) edge[bend left] node[pos=0.65]{SYN\_ACK\tss{AN}} (SYN_ACK_atob)
    (SYN_ACK_atob) edge[bend left] node[pos=0.05,yshift=-3]{SYN\_ACK\tss{NB}} (init)
    (init) edge[bend left] node[very near end]{SYN\_ACK\tss{BN}} (SYN_ACK_btoa)
    (SYN_ACK_btoa) edge[bend left] node[very near start]{SYN\_ACK\tss{NA}} (init)
    ;

    \draw[red] (SYN_btoa) .. controls +(300:2cm) and +(0:1cm) .. node[very near start]{SYN\_ACK\tss{NB}} (init);
\end{tikzpicture}
     }
    \caption{Network model under the lesspowerful MITM attack $G_{N,a}^w$}
    \label{fig:tcp:network:attack:less}
\end{figure}

\begin{figure}[htp]
    \centering
    \subcaptionbox{Peer A $G_{PA}$\label{fig:tcp:peers:timeout:A}}[\linewidth]{
        \adjustbox{scale=0.9}{
            \begin{tikzpicture}[automata, node distance=15mm]
    \renewcommand*{\pout}{\tss{AC1}}
    \renewcommand*{\pin}{\tss{C2A}}

    \node[state, initial, accepting] (closed) {closed};
    \node[state, right=of closed, ellipse] (synsent) {SYN \\ sent};
    \node[state, accepting, below=of synsent] (listen) {listen};
    \node[state, above right=of synsent] (i0) {$i_0$};
    \node[state, below right=of synsent] (i1) {$i_1$};
    \node[state, right=of listen] (i2) {$i_2$};
    \node[state, accepting, right=of i0, ellipse, minimum height=3em] (established) {established};
    \node[state, below=of established, ellipse] (synreceivced) {SYN \\ received};
    \node[state, right=of established] (i3) {$i_3$};
    \node[state, right=of i3, ellipse] (closewait) {close \\ wait};
    \node[state, above right=of closed, ellipse] (lastack) {last \\ ACK};
    \node[state, below right=of established, ellipse] (finwait1) {FIN \\ wait 1};
    \node[state, right=of finwait1, ellipse] (finwait2) {FIN \\ wait 2};
    \node[state, below=of finwait1] (i4) {$i_4$};
    \node[state, below=of finwait2] (i5) {$i_5$};
    \node[state, below left=of i4] (closing) {closing};
    \node[state, below=5mm of closing, ellipse] (timewait) {time \\ wait};

    \sffamily
    \path
    (closed) edge node{SYN\pout} (synsent)
    (synsent) edge[swap] node{SYN\_ACK\pin} (i0)
    (i0) edge node{ACK\pout} (established)
    (synsent) edge node[near start]{SYN\pin} (i1)
    (i1) edge node[near end]{ACK\pout} (synreceivced)
    (synreceivced) edge node{ACK\pin} (established)
    (closed) edge[bend right] node{listen\tss{A}} (listen)
    (listen) edge node{SYN\pin} (i2)
    (i2) edge[swap] node{SYN\_ACK\pout} (synreceivced)
    (established) edge node{FIN\pin} (i3)
    (i3) edge node{ACK\pout} (closewait)
    (closewait) edge[swap, bend right] node{FIN\pout} (lastack)
    (lastack) edge[swap, bend right] node{ACK\pin} (closed)
    (established) edge node{FIN\pout} (finwait1)
    (finwait1) edge node{ACK\pin} (finwait2)
    (finwait1) edge node{FIN\pin} (i4)
    (finwait2) edge node{FIN\pin} (i5)
    (i4) edge node{ACK\pout} (closing)
    (closing) edge[swap, bend left, in=130, out=40] node{ACK\pin} (closed)
    (i5) edge[bend left] node{ACK\pout} (timewait)
    (timewait) edge[bend left, in=125, out=40] node{deleteTCB\tss{A}} (closed)
    (listen) edge[bend right] node[right]{timeout\tss{A}} (closed)
    ;
\end{tikzpicture}
         }
    } \\ \medskip
    \subcaptionbox{Peer B $G_{PB}$\label{fig:tcp:peers:timeout:B}}[\linewidth]{
        \adjustbox{scale=0.9}{
            \begin{tikzpicture}[automata, node distance=15mm]
    \renewcommand*{\pout}{\tss{BC3}}
    \renewcommand*{\pin}{\tss{C4B}}

    \node[state, initial, accepting] (closed) {closed};
    \node[state, right=of closed, ellipse] (synsent) {SYN \\ sent};
    \node[state, accepting, below=of synsent] (listen) {listen};
    \node[state, above right=of synsent] (i0) {$i_0$};
    \node[state, below right=of synsent] (i1) {$i_1$};
    \node[state, right=of listen] (i2) {$i_2$};
    \node[state, accepting, right=of i0, ellipse, minimum height=3em] (established) {established};
    \node[state, below=of established, ellipse] (synreceivced) {SYN \\ received};
    \node[state, right=of established] (i3) {$i_3$};
    \node[state, right=of i3, ellipse] (closewait) {close \\ wait};
    \node[state, above right=of closed, ellipse] (lastack) {last \\ ACK};
    \node[state, below right=of established, ellipse] (finwait1) {FIN \\ wait 1};
    \node[state, right=of finwait1, ellipse] (finwait2) {FIN \\ wait 2};
    \node[state, below=of finwait1] (i4) {$i_4$};
    \node[state, below=of finwait2] (i5) {$i_5$};
    \node[state, below left=of i4] (closing) {closing};
    \node[state, below=5mm of closing, ellipse] (timewait) {time \\ wait};

    \sffamily
    \path
    (closed) edge node{SYN\pout} (synsent)
    (synsent) edge[swap] node{SYN\_ACK\pin} (i0)
    (i0) edge node{ACK\pout} (established)
    (synsent) edge node[near start]{SYN\pin} (i1)
    (i1) edge node[near end]{ACK\pout} (synreceivced)
    (synreceivced) edge node{ACK\pin} (established)
    (closed) edge[bend right] node{listen\tss{B}} (listen)
    (listen) edge node{SYN\pin} (i2)
    (i2) edge[swap] node{SYN\_ACK\pout} (synreceivced)
    (established) edge node{FIN\pin} (i3)
    (i3) edge node{ACK\pout} (closewait)
    (closewait) edge[swap, bend right] node{FIN\pout} (lastack)
    (lastack) edge[swap, bend right] node{ACK\pin} (closed)
    (established) edge node{FIN\pout} (finwait1)
    (finwait1) edge node{ACK\pin} (finwait2)
    (finwait1) edge node{FIN\pin} (i4)
    (finwait2) edge node{FIN\pin} (i5)
    (i4) edge node{ACK\pout} (closing)
    (closing) edge[swap, bend left, in=130, out=40] node{ACK\pin} (closed)
    (i5) edge[bend left] node{ACK\pout} (timewait)
    (timewait) edge[bend left, in=125, out=40] node{deleteTCB\tss{B}} (closed)
    (listen) edge[bend right] node[right]{timeout\tss{B}} (closed)
    ;
\end{tikzpicture}
         }
    }
    \caption{Peers with timeout}
    \label{fig:tcp:peers:timeout}
\end{figure}

    \clearpage
    \bibliography{references}
\end{document}